\documentclass[aip, amsmath,amssymb,reprint,onecolumn]{revtex4-2} 
\usepackage{amsthm,verbatim, bbm, color, graphicx, geometry, enumerate}

\usepackage{cancel}

\usepackage[utf8]{inputenc}
\usepackage[T1]{fontenc}
\usepackage{mathptmx}
\usepackage{etoolbox}

\newcommand{\ua}{\underline{a}}
\newcommand{\ub}{\underline{b}}
\newcommand{\uc}{\underline{\xi}}
\newcommand{\uu}{\underline{u}}
\newcommand{\uw}{\underline{w}}
\newcommand{\uv}{\underline{v}}
\newcommand{\ue}{\underline{e}}
\newcommand{\uz}{\underline{z}}
\newcommand{\utz}{\underline{\tilde{z}}}
\newcommand{\uzz}{\underline{0}}
\newcommand{\uk}{\underline{k}}
\newcommand{\un}{\underline{n}}
\newcommand{\um}{\underline{m}}
\newcommand{\ux}{\underline{x}}
\newcommand{\ox}{\overline{x}}
\newcommand{\oy}{\overline{y}}
\newcommand{\oz}{\overline{z}}
\newcommand{\ed}{\langle \underline{e} |}
\newcommand{\ii}{{\rm i}}
\newcommand{\ip}{{\rm i}\pi}
\newcommand{\OO}{\Omega}
\newcommand{\tO}{\tilde{\Omega}}
\newcommand{\Z}{\mathbb{Z}}
\newcommand{\Q}{\mathbb{Q}}
\newcommand{\R}{\mathbb{R}}
\newcommand{\C}{\mathbb{C}}

\newcommand{\ratthe}[2]{\vartheta \displaystyle \left[ {#1} \atop {#2}\right]}

\newcommand{\DD}{D^{(2)}_{\langle \ue_1| \, \langle \ue_2|}}
\newcommand{\DDt}{D^{(2)}_{\langle \ue_1 P^{-1}| \, \langle \ue_2 P^{-1}|}}

\newtheorem{defn}{Definition}
\makeatletter
\@ifdefinable\Olddefn{\let\Olddefn=\defn}%
\renewcommand\defn{\@ifnextchar[{\defnopt}{\Olddefn\normalfont}}%
\newcommand\defnopt[1][]{\Olddefn[{#1}]\normalfont}%
\makeatother

\newtheorem{prop}{Proposition}
\makeatletter
\@ifdefinable\Oldprop{\let\Oldprop=\prop}%
\renewcommand\prop{\@ifnextchar[{\propopt}{\Oldprop\normalfont}}%
\newcommand\propopt[1][]{\Oldprop[{#1}]\normalfont}%
\makeatother

\makeatletter
\def\@email#1#2{%
 \endgroup
 \patchcmd{\titleblock@produce}
  {\frontmatter@RRAPformat}
  {\frontmatter@RRAPformat{\produce@RRAP{*#1\href{mailto:#2}{#2}}}\frontmatter@RRAPformat}
  {}{}
}%
\makeatother

\begin{document}

\preprint{AIP/123-QED}
\title{Exploiting $\vartheta -$functions for the identification of topological materials }
\author{Emanuele Maggio} 
\email{emanuele.maggio@gmail.com}
\affiliation{Scuola Superiore Meridionale, Largo San Marcellino, 10. 80138 Napoli, Italy}

\begin{abstract}
	An exact analytical expression is derived for Bloch states in three dimensions, based on the  only assumption that the electronic wavefunction can be expanded in terms of Gaussian type orbitals. 
	The resulting expression features multidimensional $\vartheta -$functions (and their derivatives) on which the action of discrete space group symmetries is evaluated analytically and contrasted against the symmetry transformations proper of modular forms.
	We integrate group theoretical arguments with continuity requirements of the Bloch states to produce a viable algorithm for the determination of band inversions in materials with a non-trivial topological electronic band structure; the proposed methodology is then applied to two simplified materials models.
\end{abstract}
%

\maketitle

\section{Introduction}
\label{sec:Intro}

	The discovery of topological phases of matter has been one of the most consequential in recent years in condensed matter physics; originally it has been associated to the occurrence of exotic transport properties in two-dimensional systems \cite{Klitzing1980,Tong2016,Tsui1982} for which a homogeneous electron gas model \cite{Thouless1982} has provided a full account of the experimental findings.
	An ingenious dimensionality shifting algorithm \cite{Kitaev2009} has revealed the occurrence of a Bott-periodicity for the topological invariants characterising topologically non-trivial phases, thus completing their enumeration for any number of dimensions when no discrete spatial symmetries are present in the material.
	On the other hand, the description of non-trivial topological phases in three-dimensional solid-state systems has been ever reliant on the detailed study of the material's electronic structure, also in the independent electron approximation, in order to adjoin the topological invariants from the homogeneous description with additional ones able to capture the role played by the symmetries in the crystalline solid.
	Crystalline symmetries may indeed protect the topological phase \cite{Chen2013,Chiu2016}, and the development of symmetry indicators \cite{Chiu2016,Zhang2021a} has been attempted for the characterisation of crystalline topological insulators. 
	
	While the presence of symmetries complicates the theoretical analysis on one hand, on the other they enforce considerable constraints on the materials' band structure: they dictate the number of 'essential degeneracies' at high symmetry points in reciprocal space (wavevectors), and they provide a general scheme for the assignment of band connectivities based on the compatibility relations among different wavevectors. 
	The systematic study of compatibility relations among irreducible representations (Irreps) at different wavevectors, paired with combinatorial algorithms for the enumeration of all possible energy level alignments has made possible the design of high-throughput schemes \cite{Kruthoff2017,Vergniory,Tang2018a,
	Zhang2018,Marzari_WeylSMs} that would allow to systematically scan the 'chemical space' of materials candidate to host a non-trivial topological electronic structure, 
	also when time reversal is not broken: these are known as crystalline topological insulators \cite{Bansil2016, Cano2020} and represent the target system for the framework presented here. 
	Their identification can proceed by ascertaining a band inversion, that is the violation of the principle of band connectivity based on compatibility relations that preserve symmetry transformation properties for the Bloch state along the energy dispersion curve.
	In this article we aim at aiding the identification of a band inversion thanks to the requirement that the Bloch state must be a continuous function of the wavevector \cite{Bouckaert1936, Kuchment2016}, except where essential degeneracies occur. 
	This is achieved by providing an analytic expression for the Bloch states as a function of the wavevector and the real space crystallographic coordinates, which is particularly amenable since, by the same token, it fixes a gauge across the Brillouin zone \cite{Fukui2005,Fu2011}, thus troubleshooting the extra complication that comes from the discretisation of the Brillouin zone sampling in realistic \textit{ab initio} electronic structure calculations.
To conclusively determine (from a computational standpoint) the topologically nontrivial nature of the band structure once a band inversion has been identified, several approaches are available as reviewed in Ref~\onlinecite{Bansil2016}, including the evaluation of topological invariants. 
	In the present study we concentrate on the first step, the identification of a band inversion, and we leave the evaluation of topological invariants within the framework presented in this study for future research. 

\begin{figure}
	\includegraphics[width=100mm]{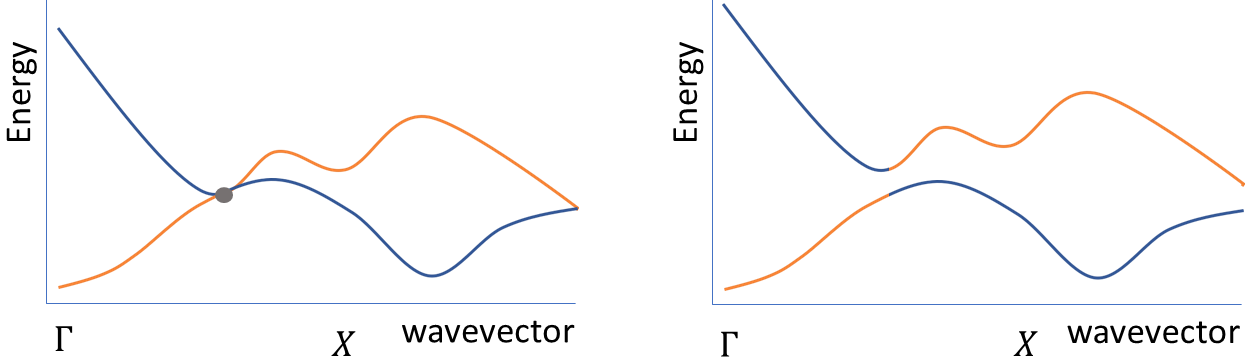}
	\caption{Schematic representation for the occurrence of a band inversion between two bands transforming like different Irreps, indicated by different colours of the energy dispersion curve. The accidental degeneracy between two states shown in the left panel is removed in the right panel, producing a band inversion.	\label{fig:cartoon}}
\end{figure}

	For a schematic textbook \cite{Jones1973} example of a band inversion consider the one dimensional Irreps $a_1$ and $a_2$ along the path in reciprocal space passing through the high symmetry line $\Delta$ with endpoints $\Gamma$ and $X$, as pictorially shown in Fig.~\ref{fig:cartoon} (left panel). 
	Assume that Irreps with the same subscript are compatible with one another and not compatible with Irreps with the other subscript.
	Continuity of the energy dispersions in reciprocal space would then imply that two bands could be identified as $\Gamma_1 - \Delta_1 - X_1$ and $\Gamma_2 - \Delta_2 - X_2$. 
	Since the associated Bloch states transform like different Irreps the two bands so identified could cross owing to the presence of an accidental degeneracy, should the energy ordering of the states at the $\Gamma -$point be the opposite of the one at the $X-$point.
	If an additional perturbation were now switched on in the Hamiltonian and the accidental degeneracy were thus to be removed, as shown in Fig.~\ref{fig:cartoon} (right panel), the continuity of the energy bands \cite{Bouckaert1936} would now imply the presence of a band inversion where the Irreps $ \Gamma_{a1}$ and $X_{a2}$, although incompatible, must be associated to the same band.
	In realistic systems this is showcased by the fact that the "symmetry eigenvalue" associated to a specific symmetry operation is not preserved by the Bloch state along the energy dispersion \cite{Bansil2016} and	the resulting band can not be labeled by a single localised descriptor.
	Localised descriptors can be identified with the elementary band decompositions (EBDs) \cite{Evarestov1997}, \textit{i.e.} the labels are  generated by the fixed points of the maximum isotropy subgroups of the space group (a. k. a. its Wyckoff positions) and the Irreps of their stabiliser that convey local symmetry information. 
	To further complicate the picture one has to bear in mind that the description in terms of localised descriptors is not unique in general \cite{Zeiner2000} and that also for a topologically trivial material elementary band decompositions can be found that do not involve a single Wyckoff position; in general, the inability to associate a single Wyckoff position as a band indicator is regarded as a sign of non-trivial topology of the band structure.
	
	On the same footing, localised Wannier functions (see Ref~\onlinecite{Marzari2012} for a comprehensive review) can be computed in the search for topological materials, to ascertain if it is possible to connect the band structure with its atomic limit.
	In the case of broken time inversion symmetry, (or for systems with non-zero Chern number) a Wannier obstruction can occur \cite{Kuchment2009, Cances2017, Auckly2018, Cornean2019b} and this is diagnostic of a topological material \cite{Cano2020}.
	If crystalline symmetries are present the straightforward computation of localised Wannier functions leads to a result that does not fulfill any symmetry constrain \cite{MLWF_Souza2001, WF_Thygesen2005} and  more sophisticated algorithms have been developed \cite{Bakhta2018} in order to include the relevant symmetries and express the resulting Wannier functions in a compact representation in terms of Gaussian type orbitals.
	In crystalline topological insulators where time inversion is preserved, a Wannier obstruction can not occur, since exponentially localised Wannier functions can always be constructed if the system is not metallic \cite{Panati2007}.
	On the other hand, the presence of a Wannier obstruction has also been reported for spinful electrons \cite{Po2018} in topologically trivial phases if crystalline symmetries are enforced, invalidating a straightforward correspondence between Wannier obstruction and band structure topology (hence \emph{fragile} topological phases need to be considered in such case) and warranting the study of alternative routes for the identification of crystalline topological insulators. 
	In this study we develop a formalism for the identification of band inversions without resorting to the calculation of Wannier functions, for systems with no spin degrees of freedom and in absence of an external magnetic field, thus preserving time inversion symmetry.
	
	Model Hamiltonians are often resorted to in the attempt to capture the salient features of the material at hand and simplify the study of  topological phases.
	Borrowing from that spirit, in this work we apply our approach to a minimal model of band structures, which is based on forming the Bloch state corresponding to a specific choice for the single particle (atomic) wavefunction, namely a Gaussian type orbital (GTO) for the spinless electron that is centred at a given Wyckoff position. 
	By constructing the Bloch function analytically we are able to explicitly evaluate its transformation properties and contrast them against the symmetry properties of the constituent GTOs. 
	In particular, this approach allows us to verify if the decomposition of the Bloch states into elementary band representations can be achieved, for a given choice of the material's electronic structure and if the compatibility relations at different values of the wavevector can be fulfilled by a given Bloch state.
	This is possible thanks to the fact that we achieve for the Bloch states a particular representation in terms of entire complex functions that are assured not to be singular for any value of the wavevector, while the wavevector changes along a path in reciprocal space.
	The approach presented here then serves as a proof of concept for more detailed \textit{ab initio} calculations of realistic materials with electronic structure codes that use GTOs as a basis set, since in that instance a consistent gauge for the Bloch functions across the Brillouin zone can be enforced with the aid of our construction.

The paper is structured as follows: in Sec.~\ref{sec:Theory} we derive the main result enabling the proposed computational method, namely the expression for the Bloch states in terms of the $\vartheta -$functions with rational characteristics and their derivatives. 
	We evaluate the action of space group operations on the Bloch states in Sec.~\ref{sec:SGaction} (derived in detail in Appendix~\ref{sec:Appx}) and we establish a link with  symmetry transformations of modular forms in Sec.~\ref{sec:TopInvModForms}.
	In the same section we also report some exact properties of the $\vartheta -$functions that can be exploited in the determination of symmetry eigenvalues and then describe the computational implementation in Sec.~\ref{sec:CompDet}. 
	Finally, we elucidate the capabilities of the proposed method for a couple of test cases considered in Sec.~\ref{sec:Expl} and we summarise our conclusions in Sec.~\ref{sec:Concl}.
	
\section{Theory}
\label{sec:Theory}

\subsection{Bloch functions from Gaussian type orbitals }
In this section Bloch functions are defined thanks to 
the Bloch-Floquet transform that allows one to decompose the Hilbert space in terms of subspaces labeled by wavevectors $\underline{\kappa}$ associated to Irreps of the group of lattice translations $\mathbb{T}$; generalising the approach in Ref. \onlinecite{Zeiner1998a}, we evaluate analytically the resulting Fourier series, thus providing a closed expression for the Bloch state stemming from a specific GTO basis function.
	Further, we evaluate the action of the material's space group symmetry operations on the analytical expressions thus derived, in order to identify basis functions for Irreps of the local symmetry group of $\underline{\kappa}$, $G_{\underline{\kappa}}$.

	A powerful consequence of the translational invariance of crystalline solid state systems is the fact that the solution to the one-body electronic problem can be factorised thanks to Bloch theorem.
	An arbitrary Bloch function can then be written as $\psi_{\underline{\kappa}} (\ux) = e^{2 \pi \ii \underline{\kappa} \cdot \ux} \Psi (\ux)$, with $\Psi (\ux + \underline{\nu}) = \Psi (\ux )$ for $\underline{\nu}$ an element of the crystal lattice $\Lambda$.
	Such a function can not be normalised unless periodic boundary conditions are enforced; when this happens one has that the resulting Bloch functions become periodic on a compact support, which need not coincide with the unit cell in general: for high symmetry wavevectors on the Brillouin zone boundary the local symmetry group might include lattice translations \cite{Bradley1972a}.
	Allowable Irreps of such symmetry groups must feature non-trivial characters for those lattice translations; we call a Bloch state a Bloch function that is periodic and transforms like an allowable Irrep of the relevant little group of the wavevector.

	To achieve a systematic decomposition of the full Hilbert space into a direct sum of mutually orthogonal subspaces $\mathcal{H}_{\underline{\kappa}}$ associated with the Irreps of the translation operators, one defines the Bloch-Floquet transform \cite{Kuchment1982, Dirl1996a}:
\begin{align}
\label{eq:ProjOp}
\mathcal{P}_{\underline{\kappa}} = \frac{1}{|BZ|} \sum_{\underline{\nu} \in \Lambda} e^{2 \pi \ii \, \underline{\kappa} \cdot \underline{\nu}} (E|\underline{\nu})
\end{align}
where the Bravais lattice $\Lambda$ is referred to the Cartesian coordinate frame and $|BZ|$ indicates the volume of the Brillouin zone.
	In order to construct the Bloch states belonging to $\mathcal{H}_{\underline{\kappa}}$, associated with a local descriptor (hence useful to identify a band inversion), we employ GTOs as basis functions for the description of the material's electronic structure.

	In Cartesian coordinates a GTO is identified by the powers $\alpha_i$ with $i = 1,2,3$ of the monomials lowering the spherical symmetry of the orbital:
\begin{align}
G^{(\beta, I)}(\ux) = C^{(\beta, I)} x^{\alpha_1} y^{\alpha_2} z^{\alpha_3} e^{-\beta \ux \cdot \ux}
\label{eq:GTO}
\end{align}
	where the composite index $I = \{ \alpha_i \}_{i =1}^3$ in the labeling of the GTO stands for the orbital type $s , p_x, p_y, p_z, d_{xy}, ...$ with the subscript corresponding to the non-zero values of the $\alpha_i$'s; for the $s$-type GTO the $\alpha$'s are all equal to zero.
	The constant $C^{(\beta, I)}$ normalises the probability density to one over the whole $\R^3$ space, that is 
\begin{align}
\left (\frac{1}{C^{(\beta, I)}} \right )^2 = \int_{\R^3} d_3 \ux |G^{(\beta, I)}(\ux)|^2
\end{align}
	and depends on the parameter $\beta$ which is related to the standard deviation of the Gaussian distribution $\sigma$ as $\beta = \frac{1}{4 \sigma^2}$  with $\sigma^2 = \int_{\R^3} d_3 \ux \, \ux^2|G^{(\beta, I)}(\ux)|^2$, and characterises the GTO itself.

	The action of lattice translations $\underline{\nu} \in \Lambda$ on GTOs is evaluated in accordance with $(E| \underline{\nu}) G^{(\beta, I)} (\ux) = G^{(\beta, I)}(\ux - \underline{\nu})\equiv G_{\underline{\nu}}^{(\beta, I)}(\ux)$ which then allows one to evaluate the Bloch state thanks to the transform in  Eq.~\ref{eq:ProjOp}:
\begin{align}
\phi_{\underline{\kappa}}^{(\beta,I)} \equiv N_{\underline{\kappa}}^{(\beta, I)} |BZ| \, \mathcal{P}_{\underline{\kappa}} G^{(\beta, I)} (\ux) = N_{\underline{\kappa}}^{(\beta, I)} \, \sum_{\underline{\nu} \in \Lambda}  e^{2 \pi \ii \, \underline{\kappa} \cdot \underline{\nu}} G_{\underline{\nu}}^{(\beta, I)}(\ux),
\end{align}
	where $N_{\underline{\kappa}}^{(\beta, I)}$ is chosen conventionally to normalise the Bloch state over the lattice's unit cell, \textit{i.e.} $\int_{u.c.} d_3 \ux \, |\phi_{\underline{\kappa}}^{(\beta, I)}(\ux)|^2 = \frac{1}{(2 \pi )^3}$.
	Generalising the one-dimensional calculation in Ref. \onlinecite{Zeiner1998}, the Fourier series above can be evaluated analytically, thus returning an exact expression for the resulting Bloch states that, in this way, is naturally associated with the GTOs' local symmetry character.

	It is convenient to consider the conventional crystal coordinate system, in which lattice translations can be represented by vectors with integer entries and hence allow to identify $\Lambda \cong \Z^3$.
	If we indicate by $T$ the linear transformation matrix such that $\ux = T \uc$, where $\uc$ are the conventional crystal coordinates, the expression for the $s$-type GTO in this coordinate system reads:
\begin{align}
G_{\underline{\nu}}^{(\beta,s)} (\ux) & = C^{(\beta,s)}
e^{-{^t(}\ux - T \un) \beta \mathbb{I} (\ux - T \un)} \nonumber \\ 
& = C^{(\beta,s)} e^{\ii \pi ^t\uc \OO \uc} \, e^{- 2 \pi \ii ^t \un \OO \uc} \, e^{\ii \pi ^t \un \OO \un } \label{eq:OO} = G_{\un}^{(\beta, s)} (\uc) 
\end{align}
in Eq.~\ref{eq:OO} the period matrix $\OO = \frac{\ii \beta}{\pi} \mathbf{G}$ has been defined, with $\mathbf{G} = {^tT} T$ the crystal metric tensor \cite{ITCA} (we indicate with ${^tT}$ the transpose of the matrix $T$).
	It is important to point out some properties of the period matrix $\OO$: it is purely imaginary and its imaginary part is symmetric and positive definite as the matrix $\mathbf{G}$ is also symmetric and positive definite by definition. 
These properties must be fulfilled in order to recast the Bloch state as a Riemann $\vartheta -$function  \cite{Mumford1983} defined in Appendix~\ref{sec:Appx}. 
	Accordingly, we have that the monomials characterising lower symmetry GTOs transform like:
\begin{align}
\langle \ue_x | \ux - T \un - T\uw \rangle & =  \langle \ue_\xi | \uc \rangle - \langle \ue_\xi | \un \rangle - \langle \ue_\xi | \uw \rangle \nonumber \\
& = f_x (\uc) - f_x (\un) - f_x(\uw) = 
f_{\langle \ue_\xi |} (\uc) -f_{\langle \ue_\xi |} (\un) -f_{\langle \ue_\xi |} (\uw) 
\label{eq:formfact}
\end{align}
where $\uw$ is the Wyckoff position where the GTO is centred and Eq.~\ref{eq:formfact} serves as a definition for two different conventions that we are going to employ when dealing with the form factors $f$, namely: the roman subscripts $x, y, \dots, i, j, \dots $ refer to the direction in cartesian coordinates, whereas, when the direction is expressed with respect to conventional coordinates, the corresponding basis vector $\langle e_\xi | = \langle e_x T |$ is used, with the subscript $\xi$ replaced by arabic numerals for GTOs with angular quantum number $\ell \geq 2$.

Finally we detail the passage to conventional crystal coordinates for the Bloch state:
\begin{align}
\phi_{\langle \underline{\kappa}|}^{(\beta, s)}(\ux) & = N^{(\beta, s)}_{\underline{\kappa}} \, \sum_{\underline{\nu} \in \Lambda} e^{2 \pi \ii \, \underline{\kappa} \cdot \underline{\nu}} \, G_{\underline{\nu}}^{(\beta, s)}(\ux) \nonumber \\
& = N^{(\beta, s)}_{\uk} \, \sum_{\un \in \Z^3} e^{2 \pi \ii \, {^t\un} \, \uk} \, G_{\un}^{(\beta, s)}(\uc) = \phi_{\langle \uk|}^{(\beta, s)}(\uc)
 \label{eq:sBloch}
\end{align}
\noindent the wavevector $\uk$ in conventional coordinates has been introduced consistently with the transformation $\underline{\kappa} T = \uk \Rightarrow {^tT} \, {^t\underline{\kappa}} = {^t \uk} $ \cite{ITCA} and the transpose superscript has been dropped in Eq.~\ref{eq:sBloch} for notational convenience, 
the original dependence on the wavevector as a row vector is however mantained in the subscript, thanks to the notation $\langle \uk |$.
	Consistently, the bra-ket notation $\langle \ua|\ub \rangle$ can also be employed to indicate the scalar product $\ua \cdot \ub \equiv {^t\ua} \, \ub$ between two vectors and we are going to use the three notations interchangeably.   
When the $s$-type GTO is centred at the Wyckoff position $\uw$ the expression for the resulting Bloch states reads:
\begin{align}
\phi_{\langle \uk|}^{(\beta,s)}( \uc|\uw, \OO) & = N_{\uk}^{(\beta,s)} \, C^{(\beta,s)} e^{\ii \pi \uc \cdot \OO \uc} \,e^{\ii \pi \uw \cdot \OO \uw} \, e^{- 2 \pi \ii \uw \cdot \OO \uc} 
\sum_{\un \in \Z^3} e^{\ii \pi \un \cdot \OO \un}  \, e^{2 \pi \ii \un \cdot (\uk - \OO \uc + \OO \uw)} \label{eq:sBloch2}
\end{align}
where the additional parametric dependence on $\OO$ has been included in Eq.~\ref{eq:sBloch2}.

Equipped with the definitions in Appendix~\ref{sec:Appx} we can further introduce a component that is common to Bloch states constructed from any type of GTO, and that it coincides with the $s$-type Bloch state, apart from its normalisation constants:
\begin{align}
\phi_{\langle \uk|}(\uc|\uw, \OO) = e^{\ii \pi \uc \cdot \OO \uc} \, e^{- 2 \pi \ii \uw \cdot \uk} \, \ratthe{\uw}{\uzz}(\uk - \OO \uc| \OO) \label{eq:phi}
\end{align}

\noindent where the $\vartheta-$function with rational characteristics has been defined in Eq.~\ref{eq:rattheta} and the complex variable $\uz = \uk - \OO \uc$ in Eq.~\ref{eq:phi} compounds information coming from the independent variables in direct and reciprocal space, also taking into account through the matrix $\OO$, details of the lattice and of the GTO. 
	The $\vartheta -$function with rational characteristics is just a translation of the original $\vartheta -$function in Eq.~\ref{eq:theta} with an additional phase factor that is automatically taken care of in Eq.~\ref{eq:phi} so that one has $\phi_{\langle \uk|} (\uc |\uw, \OO) = \phi_{\langle \uk|} (\uc - \uw |\uzz,  \OO)$, as expected since the resulting Bloch state should be independent of the choice for the unit cell origin. 
	For lower symmetry Bloch states derivatives of the $\phi -$function are necessary in order to recover the polynomials lowering the spherical symmetry, and their invariance over the unit cell origin choice  leads to the following dependence on the Wyckoff position $\uw$:
	\begin{align}
	\phi'_{\langle \ue |, \langle \uk|} (\uc |\uw, \OO) & =  2 \pi \ii \langle \ue | \uw \rangle \, \phi_{\langle\uk|} (\uc - \uw |\uzz, \OO) + \phi'_{\langle \ue |, \langle \uk|} (\uc - \uw |\uzz, \OO)
	\label{eq:phi'} \\
	\phi''_{\langle \ue_1 | \langle \ue_2 |,\langle \uk|} (\uc |\uw, \OO) & = 
	(2 \pi \ii)^2 \langle \ue_1 | \uw \rangle \langle \ue_2 | \uw \rangle \phi_{\langle\uk|} (\uc - \uw |\uzz, \OO ) + 
	2 \pi \ii \langle \ue_1 | \uw \rangle \phi'_{\langle \ue_2 |,\langle \uk|} (\uc - \uw |\uzz, \OO) + \nonumber \\
  &	2 \pi \ii \langle \ue_2 | \uw \rangle \phi'_{\langle \ue_1 |,\langle \uk|} (\uc - \uw |\uzz, \OO) +
	\phi''_{\langle \ue_1 | \langle \ue_2 | , \langle\uk|} (\uc - \uw |\uzz , \OO).
	\label{eq:phi''}
	\end{align}
	where $\phi'$ is the directional derivative in the $\langle \ue |$ direction and with respect to the complex variable $\uz = \uk - \OO \uc$  for the function defined in Eq.~\ref{eq:phi}, and similarly for the second derivative in Eq.~\ref{eq:phi''}.

	The expression for the Bloch states originating from $p$- and $d$-type GTOs is then:
\begin{align}
\phi_{\langle \ue |,\langle \uk|}^{(\beta, p_i)}( \uc|\uw, \OO) = N_{\uk}^{(\beta, p_i)} C^{(\beta, p_i)} \, \left [ f_i(\uc) \phi_{\langle \uk|}(\uc|\uw, \OO) - \frac{1}{2 \pi \ii} \phi'_{\langle \ue |,\langle \uk|} (\uc|\uw, \OO) \right] \label{eq:p_state}
\end{align}
\begin{align}
\phi_{\langle \ue_1 | \langle \ue_2|, \langle \uk|}^{(\beta, d_{ij})}( \uc |\uw, \OO) = N_{\uk}^{(\beta, d_{ij})} C^{(\beta, d_{ij})} &\left [ f_{\langle \ue_1|}(\uc) f_{\langle \ue_2|}(\uc) \phi_{\langle\uk|}(\uc|\uw, \OO) - \frac{1}{2 \pi \ii} f_{\langle \ue_1 |}(\uc) \phi'_{\langle \ue_2|, \langle \uk|} (\uc|\uw, \OO) \right . \nonumber \\
& \left . - \frac{1}{2 \pi \ii} f_{\langle \ue_2 |}(\uc) \phi'_{\langle \ue_1|, \langle \uk|} (\uc|\uw, \OO) + \frac{1}{(2 \pi \ii)^2} \phi''_{\langle \ue_1| \langle \ue_2|, \langle \uk|} ( \uc|\uw, \OO)\right ] \label{eq:d_state}
\end{align}
which are the minimal expressions that allow the resulting Bloch state to be invariant with respect to the origin choice for the unit cell.
	The use of $\vartheta-$functions with rational characteristics also incorporates the invariance of the Bloch state with respect to integer translations of $\Re\{\uz \}=\uk$: this property is reported in Proposition~\ref{prop:quasiperiod}-1. in Appendix~\ref{sec:Appx}.
	Owing to this invariance, the case of a non-zero magnetic field can not be described in the present framework.

\subsection{Transformation under crystalline symmetries}
\label{sec:SGaction}
	The action of space group symmetry operations can be evaluated analytically on the Bloch states so expressed; for a space group operation $F = (P|\uu)$, $P$ represents the (im)proper rotation matrix and $\uu$ can be either the a translation by a fraction of the lattice vector for non-symmorphic space groups or a centering vector for non-primitive unit cells, or a  translation by a lattice vector when required by the local symmetry group of the wavevector $G_{\uk}$. We then have:
\begin{align}
F \left [ \phi_{\cdot, \langle \uk |}^{(\beta, I)} ( \uc|\uw, \OO)\right] 
 = \phi_{\cdot , \langle  \uk P^{-1} |}^{(\beta, I)}( \uc |P\uw + \uu, \tO) , \label{eq:Transf_orb}
\end{align}
where the placeholder subscript on the left hand side stands for $\langle \ue |$ for $I=p_i$, $\langle \ue_1 | \langle \ue_2 |$ for $I = d_{ij}$ and it is left out for $I=s$. 
	The corresponding transformed quantities on the right hand side read as $\langle \ue P^{-1}|$ for $I=p_i$ and $\langle \ue_1 P^{-1} | \langle \ue_2 P^{-1} |$ for $I=d_{ij}$.
	The action of a space group operation on a Bloch state made up by a GTO shifts the state to a new Wyckoff position and transforms the period matrix by a similarity transformation as $\tO = {^t P}^{-1} \OO P^{-1}$ (more on this action below), whereas the row vectors $\langle \ue|$, $\langle \uk |$ transform according to the contragradient representation of the group; the derivation of Eq.~\ref{eq:Transf_orb} above is reported in Appendix~\ref{sec:Appx} for the various cases $I = s, p, d$.
	A key advantage of using the $\vartheta -$functions with rational characteristics instead of the ordinary Riemann $\vartheta -$function is that when $\uu = \un$ happens to be an integer lattice translation, the nontrivial transformation of the Bloch state is automatically enforced by Proposition \ref{prop:quasiperiod}-3.
	In this context that statement reads $\ratthe{\uw + \un}{\uzz}(\uz | \OO) = \ratthe{\uw}{\uzz}(\uz|\OO)$ which then allows to easily evaluate the translated Bloch state as: $\phi_{\cdot, \langle \uk|}^{(\beta, \cdot)}( \uc|\uw + \un,\OO) = e^{-2\pi \ii \un \cdot \uk} \, \phi_{\cdot, \langle \uk|}^{(\beta, \cdot)}(\uc|\uw, \OO)$. 
	Further, the invariance of the $\vartheta -$functions with respect to integer lattice translations correctly reflects the fact that the physical information for the construction of the Bloch state is only contained in the Wyckoff position where the orbital is centred and not in the particular unit cell in the crystal; for this reason the Wyckoff position can be always chosen to belong to the vector space $\R^3 / \Z^3$, consistently with the affine embedding theorem \cite{Eick2006}.

	To determine the character of a given Bloch state, one can proceed by systematically applying the transformations reported in Eq.~\ref{eq:Transf_orb} and comparing the transformed states (averaged over conjugacy classes) with the characters of $G_{\uk} $.
	For the sake of concreteness in the rest of this study we are going to consider two space groups with a primitive cubic lattice system. 
	In such a situation the transformed  period matrix  $\tO = \OO$ for all symmetry elements in the space group since the metric tensor is a scalar matrix and the rotational matrices are orthogonal.
	In this case the action of an element $F \in G_{\uk}$ on an $s$-type GTO based Bloch state can be concisely evaluated as follows:
	\begin{align}
	\frac{F[\phi]}{\phi} = e^{-2 \pi \ii (P \uw - \uw + \uu) \cdot \uk} \, \,  \frac{\ratthe{P \uw + \uu}{\uzz}(\uz | \OO)}{\ratthe{\uw}{\uzz}(\uz | \OO)}
	\label{eq:ratAction}
	\end{align}
and the ratio of $\vartheta -$functions with rational characteristics has a direct connection with the Irrep's characters.
In the case of a simple band (\textit{i.e.} a one-dimensional Irrep) it  coincides with said character, in agreement with the previous literature (see for example Equation 6 in Ref. \onlinecite{Michel1992}).
	When a composite band is considered a linear combination of GTOs is required to form the appropriate Bloch state, as the space of symmetry abiding partner functions has dimension bigger than one.
	Symmetry operations that are not self-conjugate in this case can map one partner function to the other, making the numerical comparison inconclusive; on the other hand if the element $F$ is varied over its conjugacy class and the left hand side is summed over such elements, for a Bloch state that transforms like a selected  Irrep the ratio on the right hand side above evaluates to $\chi_i |\mathfrak{Cl}_i|/N_{orb}$, where $\chi_i$ is the Irrep character for the $i$-th conjugacy class of size $|\mathfrak{Cl}_i|$ and $N_{orb}$ is the number of GTOs.
	Note that Equation (\ref{eq:ratAction}) generalises the expression reported in the previous literature to include the action of fractional translations.

\subsection{Transformations as modular forms}
\label{sec:TopInvModForms}
	Now we are in a position to examine the full bearing of the mathematical properties of $\vartheta -$functions on the evaluation of symmetry indicators.
	First, we examine the action of the symmetry group for modular forms and we put it in correspondence with the action of space group elements that we investigated in the previous section.
	Then we specify the discussion to the case of the inversion symmetry: it turns out that there is a one-to-one correspondence between the sets $\Xi_s$ of points $ \uz \in \C^g$ ($g = 3 $ in our case) for which the $\vartheta -$function with rational characteristics is equal to zero and the pairs $s = (\uw, \uk)$ that have half-integer coordinates. This theorem was stated in Ref.  \onlinecite{Mumford1983} and it is reported in Appendix~\ref{sec:AppxThm} for completeness. 
	We derive an expression for the inversion symmetry eigenvalue -indicated with the symbol $e_*(s)$- in terms of those coordinates and we link this expression to  a (previously reported \cite{Michel1992,Zak1991,Atala2013}) difference in Berry phase.
	The conclusive remark in this section is a corollary which associates the value of $e_*(s) = \pm 1$ to whether the complex vector formed from $s$ is a zero of the $\vartheta -$function.
	The converse of these facts will be used in Sec.~\ref{sec:SG212} to characterise the occurrence of a band inversion:
	when the crystal is not centrosymmetric, the set  $\Xi_s$ can be conserved  along a path in reciprocal space, since different wavevectors can be associated to the same set $\Xi_s$, thus highlighting what could be a new way to characterise the occurrence of a band inversion.
	
	In the general case when the action of a space group operation returns a transformed $\tO \neq \OO$, one has to consider the action of the symplectic group $Sp(2g, \R)$ on the period matrix. 
	The action of 
	$\gamma =\begin{pmatrix}
	A  & B \\
	C  & D 	
	\end{pmatrix} \in Sp(2g, \R)$ on $\OO$ returns $ \OO' = (A \OO + B ) (C \OO + D)^{-1}$ (see Ref. \onlinecite{Diamond2005} on sec. 4.9 and Ref. \onlinecite{Mumford1983} on page 173), so it is sufficient to consider one generator 
	$\gamma' =\begin{pmatrix}
	A  & 0 \\
	0  & {^t A^{-1}}
	\end{pmatrix} $ of $Sp(2g, \R)$ and set $A = {^tP^{-1}}$ to establish a connection between the action of space group operations on a Bloch state and the transformation properties of $\vartheta -$functions as modular forms; the action of the other generator of $Sp(2g, \R)$ is to endow the matrix $\OO$ with a symmetric real part and it is not considered in this article.
	The fundamental functional equation for the $\vartheta -$function as a modular form is the following:
	\begin{align}
	\gamma [\vartheta(\uz | \OO)] = \vartheta({^t(} C \OO + D)^{-1} \uz | \OO') = \zeta_\gamma \, (\det(C \OO + D))^{\frac{1}{2}} \, e^{\ii \pi \uz \cdot (C \OO + D)^{-1} \, C \, \uz } \, \vartheta (\uz | \OO)
	\label{eq:modularEq}
	\end{align}
	where $\zeta_\gamma$ is an eighth root of unity that depends on the symmetry operation only.
	If we specify $\gamma = \gamma'$ as above and consider the translate of the variable $\uz = - \OO \uc + \OO \uw + \uk$ then the modular equation \ref{eq:modularEq} takes the more yielding form:
	\begin{align*}
	\vartheta ({^t P}^{-1} (- \OO \uc + \OO \uw + \uk) | {^tP}^{-1} \OO P^{-1}) = \zeta_{\gamma'} \sqrt{\det (P)} \, \vartheta (- \OO \uc + \OO \uw + \uk | \OO)
	\end{align*}
	or, if both the Wyckoff position and the wavevector are rational vectors,
	\begin{align}
	\ratthe{P\uw}{{^tP}^{-1} \uk} (- \tO P \uc | \tO) = \zeta_{\gamma'} \sqrt{\det (P)} \ratthe{\uw}{\uk}(-\OO \uc | \OO)
	\end{align}
	which allows immediately to determine the action of a point group operation on the $\vartheta -$function with rational characteristics:
	\begin{align}
	P \left[ \ratthe{\uw}{\uk}(-\OO \uc | \OO) \right] = \ratthe{\uw}{\uk}(-\OO P^{-1} \uc | \OO) = \zeta_{\gamma'}^{-1} (\det (P))^{-\frac{1}{2}} \ratthe{P\uw}{{^tP}^{-1} \uk}(- \tO \uc | \tO).
	\label{eq:ratthePoint}
	\end{align}
	By comparing Eq.~\ref{eq:ratthePoint} with the transformation under the action of a space group operation in Eq.~\ref{eq:rattheTransf}, it follows that $\zeta_{\gamma'}^{-1} = \sqrt{\det (P)}$, thus allowing us to track down an otherwise elusive root of unity.
	
	If we specify the discussion above to the inversion symmetry by setting $P = - \mathbb{I}$, we have that
	for the ordinary $\vartheta -$function this action is actually trivial, that is $\vartheta(\uz | \OO) = \vartheta(-\uz | \OO)$ which follows from the definition in Eq.~\ref{eq:theta} and a re-indexing of the summation therein.
	On the other hand, in the case of a function with rational characteristics we have:
	\begin{align*}
	\ratthe{\uw}{\uk}(- \uz | \OO) & = 
	\sum_{\un \in \Z^g} e^{\ip (\un + \uw)\cdot \OO(\un + \uw)} \, e^{2\ip (\un + \uw) \cdot (-\uz + \uk)} \\
	& =	\sum_{\un \in \Z^g} e^{\ip (-\un - \uw)\cdot \OO(-\un - \uw)} \, e^{2\ip (-\un - \uw) \cdot (\uz - \uk)}	=	
	\ratthe{-\uw}{-\uk}(\uz | \OO) 
	\end{align*}
	since $\un$ and $-\un$ span the same lattice.
	To make contact with the assumptions of the theorem in Appendix~\ref{sec:AppxThm}, we assume that both $\uw $ and $\uk$ have half-integer entries, then we can write
	\begin{align*}
	\ratthe{-\uw}{-\uk}(\uz | \OO) = 			\ratthe{\uw -2\uw }{\uk-2\uk}(\uz | \OO) = e^{-4 \ip \uw \cdot \uk} \ratthe{\uw}{\uk}(\uz | \OO)
	\end{align*}
	 which follows from Proposition \ref{prop:quasiperiod}-3.,
	 summarising the derivation above we can evaluate the parity eigenvalue as:
	\begin{align}
	\ratthe{\uw}{\uk}(-\uz | \OO) & = e_*(\uw,\uk)\, \ratthe{\uw}{\uk}(\uz | \OO), \nonumber \\
	e_*(\uw, \uk) & = (-1)^{4 \, \uw \cdot \uk}
	\label{eq:parity}
	\end{align}
	that takes values in $\Z /2\Z$, and thus can serve as a simple \textit{rule of thumb} to help determine the parity of a given Bloch state.
		For Bloch states that are made up by GTOs with angular quantum number $\ell \neq 0$ the additional factor $(-1)^\ell $ crops up when evaluating the parity transformation, this term being constant across the Brillouin zone, it bears no consequence for defining the resulting parity eigenvalue.
	We point out that the exponent in Eq.~\ref{eq:parity} is exactly the difference in Berry phase evaluated in Ref. \onlinecite{Michel1992}, and its role in characterising the topological nature of the band considered is going to  be conclusively established once the calculation of topological invariants will be implemented in the present framework.
	
	However, at present we can make a remark concerning the correspondence between the value of the parity eigenvalue in Eq.~\ref{eq:parity} and the zeroes of the corresponding $\vartheta -$function given in Ref. \onlinecite{Mumford1983}. 
	Given $s=(\uw, \uk) \in \frac{1}{2} \Z^{2g}/\Z^{2g}$, the theorem in Appendix~\ref{sec:AppxThm} states that there is a one-to-one correspondence between such a vector $s$, and the zeroes of the Bloch state, indicated with $\Xi_s$.
	When the space group does not contain the inversion symmetry, such correspondence no longer holds and one has that the same set of zeroes can be shared among different $s$ vectors, which implies that the same set $\Xi_s$ can characterise Bloch states at different wavevectors and/or associated with different Wyckoff positions.
	We are going to showcase this claim in Sec.~\ref{sec:SG212} where the occurrence of a band inversion for SrSi$_{\textrm{2}}$ would lead to the set $\Xi_s$ being preserved as the wavevector is varied along the energy dispersion curve.

\section{Computational implementation}
\label{sec:CompDet}
The numerical evaluation of the $\vartheta -$functions and their derivatives has been carried out in \textsc{Matlab} \cite{Matlab2024}  using the pointwise approximation reported in Ref. \onlinecite{Deconinck2003}.
	The numerical error has been set in as an input parameter to the value $10^{-8}$ and the resulting error on the computed Bloch states has been estimated by looking at sections of the unit cell where nodal planes would be present: a numerical noise of the order of $10^{-17}$ would usually be observed, which we deemed sufficient for an accurate estimation of the symmetry character of the resulting Bloch state under the action of symmetry operations.
	
	For a given wavevector, we can compute the local symmetry group $G_{\uk}$ as specified in our previous work \cite{Maggio2023} and the action of the symmetry operations leads to the transformed Bloch states evaluated according to Eq.~\ref{eq:Transf_orb}.
	To expedite the numerical evaluation of the symmetry transformation, a point at random in the unit cell is chosen, so to form the complex variable $\uz = \uk - \OO \uc$; the original Bloch states are not normalised, that is the constant $N_{\uk}^{(\beta, \cdot)}$ is set equal to 1, whereas the Gaussian orbitals are normalised once and for all at the start of the calculation.
	The average over the conjugacy class elements is then computed and the result is then normalised by the value of the original Bloch state, in other words, the ratio on the left hand side of Eq.~\ref{eq:ratAction} is evaluated for the given Bloch state and the result is then compared with the Irreps characters computed as described in Ref. \onlinecite{Maggio2023}.

	Independently of the action of the space group elements on the Bloch states, we obtain the EBD for the space group at hand: this standard group theoretical calculation has been detailed in the previous literature (see for instance Ref. \onlinecite{Evarestov1997}) and it consists in inducing the space group representations for each high symmetry wavevector and then finding the components of the Wyckoff position's Irreps subduced by the Irreps of the space group.
	\begin{table}
	\caption{\label{tab:EBD207} Elementary Band decomposition for space group $P432$ no. 207 and Wyckoff position $\uw_a = [0, 0, 0]$. The character tables for the local symmetry groups of the wavevectors listed and $\mathsf{Stab}(\uw_a)$ are reported in Appendix~\ref{sec:AppxCC}.}
	\begin{tabular}{@{}lllll}
	\hline
	 & $\Gamma$ & $X$ & $M$ & $R$ \\
	\hline
	$(\uw_a, a_1)$ & $\Gamma_{a1}$ & $X_{a8}$ & $M_{a8}$ & $R_{a3}$\\ 
	$(\uw_a, a_2)$ & $\Gamma_{a2}$ & $X_{a4}$ & $M_{a2}$ & $R_{a2}$\\ 
	$(\uw_a, e_3)$ & $\Gamma_{e3}$ & $X_{a4} \bigoplus X_{a8}$ & $M_{a2} \bigoplus M_{a8}$ & $R_{e5}$\\
	$(\uw_a, t_4)$ & $\Gamma_{t4}$ & $X_{a6} \bigoplus X_{e9}$ & $M_{a6} \bigoplus M_{e9}$ & $R_{t8}$\\
 $(\uw_a, t_5)$ & $\Gamma_{t5}$ & $X_{a2} \bigoplus X_{e9}$ & $M_{a4} \bigoplus M_{e9}$ & $R_{t7}$\\
 	\hline
	\end{tabular}
	\end{table}
	We tabulate the results as shown in Tab.~\ref{tab:EBD207}: each row is labelled by the Wyckoff position $\uw$ considered and the corresponding Irrep of $\mathsf{Stab}(\uw)$, its stabiliser, also known as the isotropy group.
	Each column is indexed by a label that contains the wavevector, a letter that conventionally reflects the Irrep degree ($a$ being one dimensional, $e$ two dimensional, $t$ three dimensional and so on) and a digit that refers to the row entry in the character table for $G_{\uk}$, reported in Tabs.~\ref{tbl:StabWa_SG207}-\ref{tbl:G_R212} in Appendix~\ref{sec:AppxCC}.
	The table entries in Tabs.~\ref{tab:EBD207}-\ref{tab:EBD212} report the decomposition frequencies of the Irreps so described; Irreps associated with different $G_{\uk}$ that decompose according to the same Irrep of $\mathsf{Stab}(\uw)$ are said compatible.
	With a little abuse of notation we are going to say that an Irrep of $G_{\uk}$ is compatible with $( \uw, i)$ if it has a non-zero decomposition frequency.
	
	A sketch of the present algorithm is then as follows: 
	first we find GTOs (or linear combinations thereof) -centred in the appropriate Wyckoff position- that form a basis for a given Irrep of $\mathsf{Stab}(\uw)$; this means that they remain invariant when acted on by the chosen Irrep.
	If the $i$-th Irrep (of degree $d_i$) affords the matrix representation $\mathfrak{M}^{(i)}$, and if we indicate the action of the element $g \in \mathsf{Stab}(\uw)$ on each partner function $\psi_k$ as $g[\psi_k]$ we then have that the resulting symmetrised function is given by \cite{Kim1999}:
	\begin{align}
	\overline{\psi}_j = \frac{d_i}{|\mathsf{Stab}(\uw)|} \sum_{k = 1}^{d_i} \sum_{g \in \mathsf{Stab}(\uw)} \mathfrak{M}_{kj}^{(i)}(g) \, g[\psi_k]
	\label{eq:BasisStab}
	\end{align}
	The algorithm identifies a basis for the representation when we find $\overline{\psi} = \psi$.
	
	From the GTOs so selected, the corresponding Bloch states can be analytically computed with the methods in Sec.~\ref{sec:Theory} and their characters can be evaluated as reported above. 	
	In addition, the present scheme can also be applied to the case of a composite band: this is an electron dispersion where more branches are labelled by the same local descriptor $(\uw, i)$ (also known as a continuity chord), and which also can present the occurrence of energy level crossing as shown in Sec.~\ref{sec:SG207}.
	Next we are going to consider two space groups as test cases where the previous algorithm can be put to fruition.
	
\section{Examples}
\label{sec:Expl}
	In this section we apply the algorithm outlined in the preceding section to two space groups: we consider no. 207 which is a symmorphic space group belonging to  the octahedral holohedry and no. 212 which belongs to the same arithmetic crystal class as the previous one, the latter is of relevance for the study of topological phases since it is the space group of  SrSi$_\textrm{2}$, a candidate topological material  \cite{Singh2018}. 
	As we are interested in the general features of the method proposed, we forgo the attempt to include the materials' details in our description, as such we set the lattice constant equal to 1 and the broadening $\beta = $ 0.1 in this section. 

\subsection{Space group no. 207 (P432)}
\label{sec:SG207}
\begin{figure}
	\includegraphics[width=140mm]{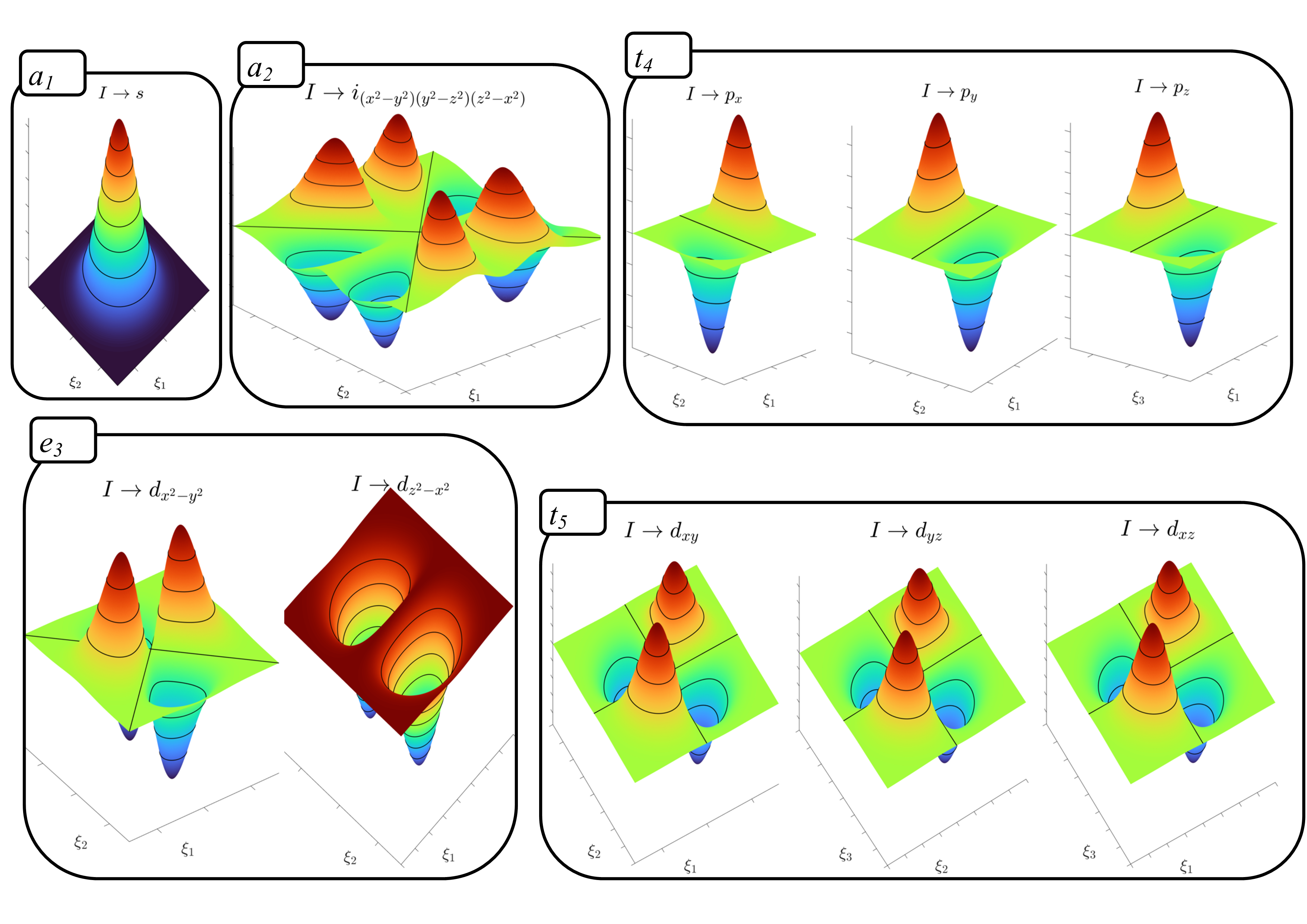}
	\caption{Gaussian type orbitals transforming like Irreps of $\mathsf{Stab}(\uw_a)$ for SG no. 207, the orbitals are plotted along the $\xi_3 =0$ plane, except when this would be a nodal plane for the function shown.
The higher angular momentum $i$ GTO has been approximated by taking the product of the relevant $d$-type GTOs.	\label{fig:GStabWa_207}}
\end{figure}

\begin{figure}
\includegraphics[width=160mm]{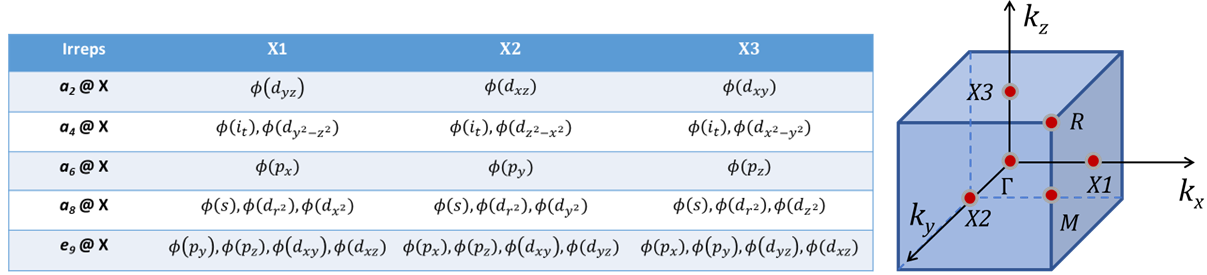}
\caption{(Left panel) Bloch states transforming like Irreps on the leftmost column formed using GTOs indicated in brackets for different choices of the coordinates representatives of the $\uk-$point $X$. The shorthand notation for the GTOs is self-explanatory except for the case $i_t \equiv i_{(x^2 - y^2)(y^2 - z^2)(z^2 - x^2)}$ and $d_{r^2} \equiv d_{x^2 + y^2 + z^2}$. (Right panel) Schematic of the first Brillouin zone highlighting the high symmetry $\uk-$points coordinates.
\label{fig:X_BlochStates}}
\end{figure}

	Starting with SG no. 207 we identify the GTOs forming a basis for the local symmetry indicators of the Wyckoff position $\uw_a = [0 , 0 , 0]$: these are shown in Fig.~\ref{fig:GStabWa_207}, that means that they are invariant under the action of $\mathsf{Stab}(\uw_a)$ as defined in Eq.~\ref{eq:BasisStab}.
	It is worth pointing out that for the two dimensional manifold of type $e_3$ the $d$-type Gaussian $z^2 - x^2$ has been selected instead of the radially symmetric (in the $\xi_3 = 0$ plane) $3z^2 - r^2$ that is usually employed as an atomic orbital, even though the latter would also have the required symmetry.
	However, the corresponding Bloch states would then fail to transform like any Irrep of $G_{\uk}$ for wavevectors other than $\Gamma$, thus making that Bloch state not compatible for lower symmetry wavevectors, as showcased in Fig.~\ref{fig:X_BlochStates} for the (symmetrically equivalent) $X-$points.
	What can easily be observed is that as the symmetry is lowered by going from the $\Gamma$- to the $X-$point, degeneracies among, let's say, $p$-type Bloch states are lifted in accordance with the EBD in Tab.~\ref{tab:EBD207}: they decompose according to $\Gamma_{t4} \approx X_{a6} \bigoplus X_{e9} $ but how that manifold splits among the lower symmetry ones depends on the choice made for the actual $X-$point coordinates, as this will impact on the elements that form the group $G_X$. 
	In fact for equivalent $\uk$-points we have that the respective $G_{\uk}$ are conjugate groups in $G$, but they are by no means the same group.
	
\begin{figure}
\includegraphics[width=140mm]{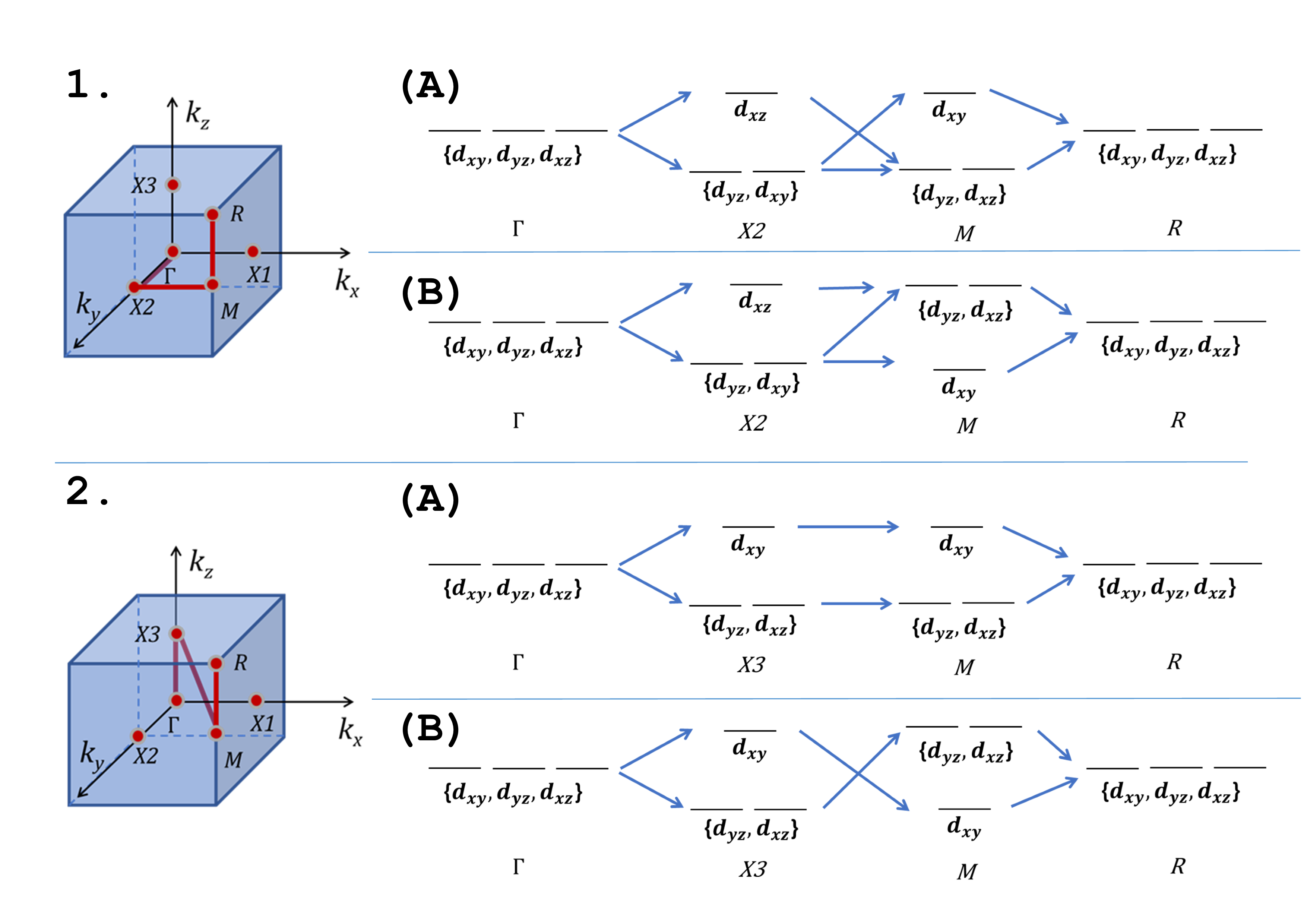}
\caption{(Left panels) Paths in reciprocal space along which the wavevector is varied. (Right panels) Make up of the Bloch states labelled by the local symmetry indicator $(\uw_a, t_5)$ as a function of the $\uk-$point coordinates for the two energy level alignments $(\mathtt{A})$ and $(\mathtt{B})$.
\label{fig:DispRels_SG207}}
\end{figure}

	This fact substantiates the analysis reported in Fig.~\ref{fig:DispRels_SG207}, where two equivalent paths in reciprocal space are contrasted.
	The paths are shown on the left panel in Fig.~\ref{fig:DispRels_SG207}: they include respectively the points $\{\Gamma - X2 - M - R \}$ in panel 1 (top) and the points $\{\Gamma - X3 - M - R \}$ in panel 2 (bottom); for each path the Bloch states formed from the $\{ d_{xy}, \, d_{yz}, \, d_{xz}\}$ GTOs are constructed and their symmetry evaluated. 
	Two different energy orderings of the states are considered and indicated by letters ($\texttt{A}$) and ($\texttt{B}$), and the connectivity between energy levels at different $\uk$-points is assigned based on the local label for the GTO from which the Bloch state is formed, this would not have been possible by using the information contained in the EBD, since the energy levels involved form a single, composite energy band.
	What we observe is that for either choice ($\texttt{A}$ or $\texttt{B}$) of the energy orderings we have the occurrence of an energy level crossing along one of the paths considered. 
	The presence this symmetry enforced level crossings is a robust feature of the system given that the same energy level alignment must be preserved along the two paths in reciprocal space, assuming the Hamiltonian invariance under space group operations.
	We maintain that the analysis showcased here could be employed systematically for the identification of band crossings also when the energy levels are not associated with the same localised descriptor.

\subsection{Space group no. 212 (P4$_\textrm{3}$32)}
\label{sec:SG212}

	Next we consider a non-symmorphic space group belonging to the same arithmetic crystal class of the previous example: space group no. 212 has been chosen since this is the symmetry group of SrSi$_\textrm{2}$, which has been identified as a candidate topological material in Ref. \onlinecite{Singh2018}.
	This system represents an intriguing test case since successive thermoelectric transport calculations have failed to identify topological contributions to transport and have highlighted how the band structure crucially depends on the lattice constant value employed \cite{Shiojiri2021}, thus providing competing experimental findings. 
	On the other hand, the electronic structure of the valence and conduction bands is straightforward, the first comprising mostly of  Si atoms' $p$-orbitals and $d$-GTOs on Sr atoms making up the most of the latter \cite{Chen2011a}, thus making it suitable for a simplified model representation in the spirit of the approach we propose.
	
	The requirement of a Bloch state to be analytical along the wavevector dispersion is violated only for a discrete set of points in reciprocal space where the multiplicity of the energy band changes, because for a degenerate manifold any linear combination of the partner functions can form a Bloch state with the required symmetry.
	When accidental degeneracies are present, the multiplicity of the manifold does not change, only two states pertaining to different bands happen to coincide. 
	Continuity of the Bloch states then can help identify the interpolating function that is required in order to make the two distinct states  coincide at the wavevector where the band crossing occurs.
	Consistently with the discussion in Sec.~\ref{sec:Intro}, the interpolating function that we will identify will not be associated to a single Wyckoff position.
	The aim of this study is not to provide a definitive answer to the alternative explanations for the experimental observations that have been put forward: this would not be  possible for a symmetry based analysis that can not access the energy scales involved, but rather to expound the possibility of such a band crossing and to relate its occurrence to the principles exposed in Sec.~\ref{sec:TopInvModForms}.
	
\begin{figure}
\includegraphics[width=140mm]{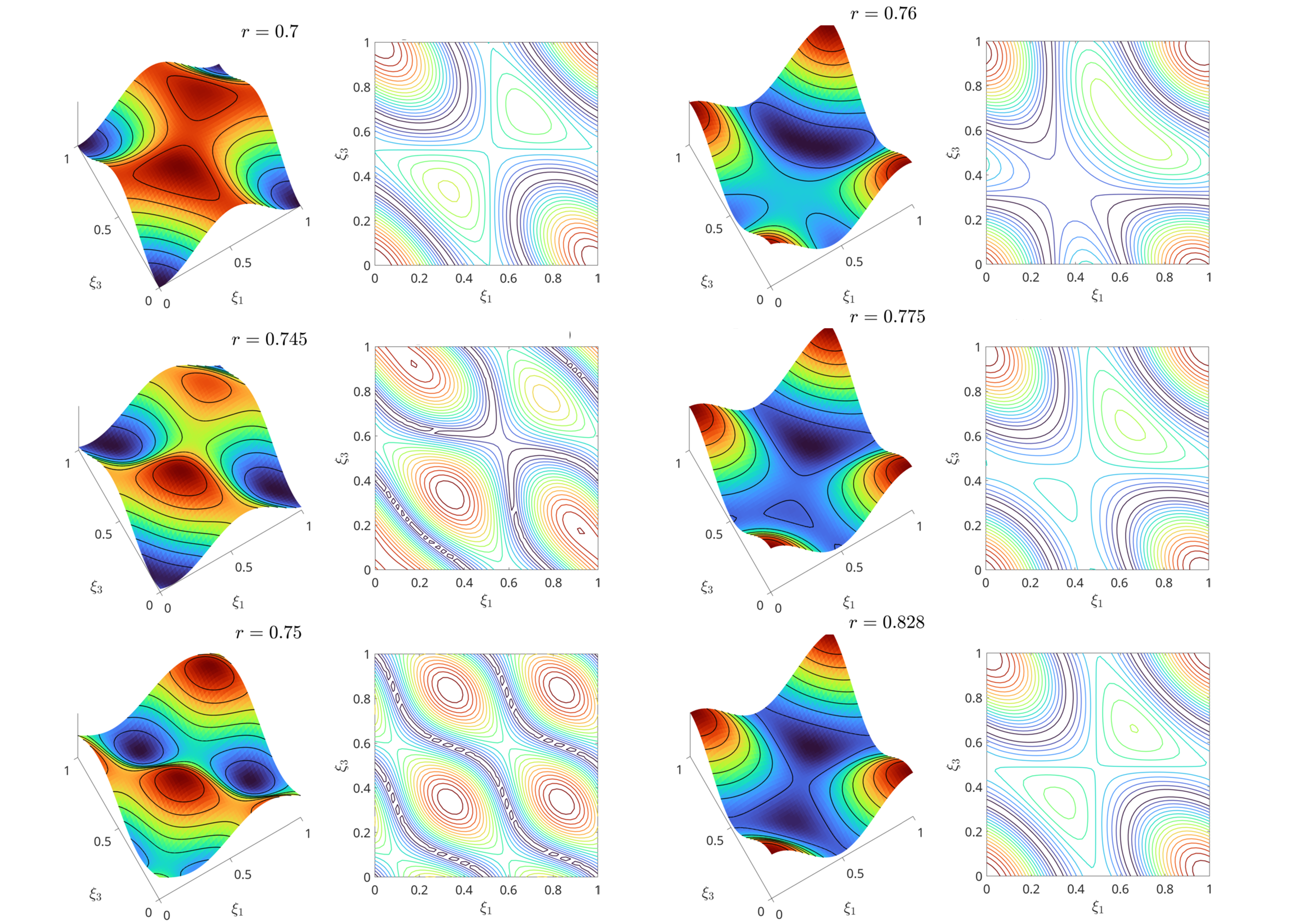}
\caption{\label{fig:BS_Siposn} Bloch states at the $\Gamma$-point originating from a symmetric superposition of $p-$type GTOs centred at Wyckoff position $\uw_c =[r,r,r]$. For each value of $r$ reported we are showing the real part (left) and the absolute value (right) of the Bloch state in the plane $\xi_2 = 0$.}
\end{figure}

	It has been pointed out already that different atomic orbitals contribute to the valence and conduction bands, as such, different localised descriptors can be assigned to the two bands we consider: the Sr atoms occupy the Wyckoff position $\uw_a$ with multiplicity four and coordinates in the orbit of $[\frac{\textrm{1}}{\textrm{8}},\frac{\textrm{1}}{\textrm{8}},\frac{\textrm{1}}{\textrm{8}}]$, while Si atoms are in position $\uw_c$ which has as representative coordinate the axis $[r, r, r]$.
	The corresponding Bloch states are significantly affected by the parameter $r$ determining the Si atom positions; this is shown in Fig.~\ref{fig:BS_Siposn} where the values of $r$ are varied up to the reference value reported in Ref. \onlinecite{Brutti2006}.
	Since the Wyckoff position comprises of a continuous manifold, the $t_5$ symmetry of the Bloch state shown in Fig.~\ref{fig:BS_Siposn} is not broken as the Si atoms position is varied, yet from a qualitative standpoint it looks that the state observed for $r=$0.7 or $r=$0.828 (they differ only by an overall sign) undergoes a significant rearrangement with an intermediate state being reached for the value of the fractional coordinate $r=$0.75.
	In the rest of our discussion we specify the coordinates in $\uw_c$ in agreement with previous reference to the value $r=$0.828.

\begin{figure}
\includegraphics[width=140mm]{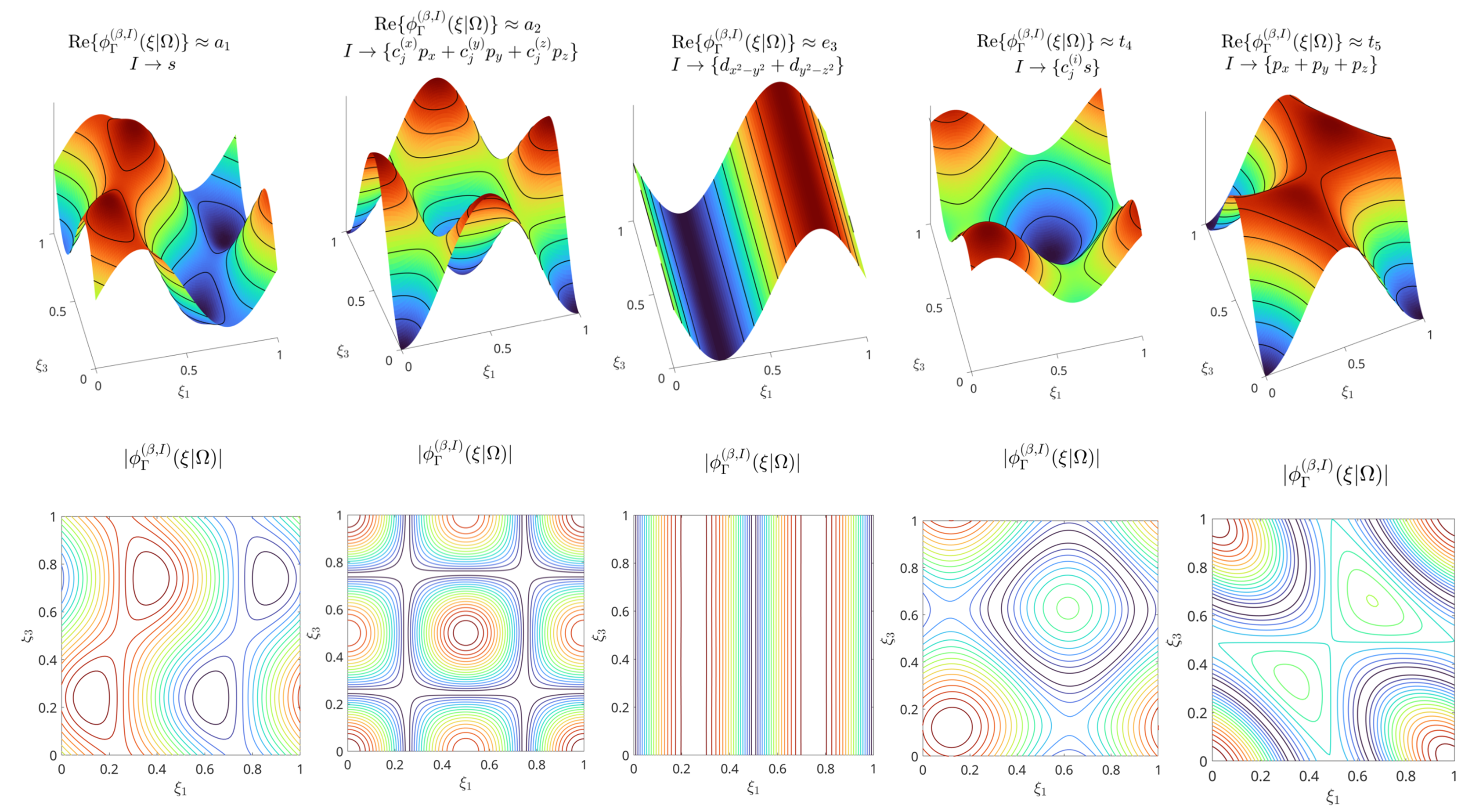}
\caption{(Top panels) Real part of the Bloch states transforming like Irreps of $G_\Gamma$ for space group 212. (Bottom panels) Absolute value for the states above. The generating GTOs (indicated by the values of the composite index $I$) are centered on the coordinates in the orbit of $\uw_a$.
\label{fig:BS_GammaSG212}}
\end{figure}

	We set out our analysis by considering the factor group $G/\mathbb{T}$ which returns the local symmetry group $G_\Gamma \cong \mathsf{Sym}(4)$, thus at the $\Gamma$ point we have five Irreps (reported in Tab.~\ref{tbl:G_Gamma212}) with corresponding Bloch states being shown in Fig.~\ref{fig:BS_GammaSG212} constructed from GTOs in position $\uw_a$.  
	Since the orbit of $\uw_a$ contains more than one coordinate, the Bloch states are built from the superposition of GTOs centred at each  coordinate in the orbit \cite{Zeiner2000}.
	The coefficients of such linear combinations are understood to be all equal to one when not  explicitly included in the corresponding heading in Fig.~\ref{fig:BS_GammaSG212}: this is the case for the $\Gamma_{a1}$ and $\Gamma_{t5}$ states, made up respectively by a symmetric superposition of $s$ and $p_x, p_y, p_z$ orbitals, and for the $\Gamma_{e3}$ state which is spanned by any antisymmetric pairwise combination of $d$-type orbitals featuring maximal degree monomials.
	A basis for the remaining Irreps can be obtained by considering the coset representatives $\{j=1 \dots 4 \}$ for the quotient $G_{\Gamma} / \mathsf{Stab}(\uw_a) \cong \mathsf{Sym}(4) / \mathsf{Sym}(3)$, specifically, the coefficients $\{c^{(i)}_j \}_{i=1}^3$ of the linear combination are given by the entries of the (3 dimensional) matrix representations of the corresponding elements in $G_\Gamma$.

	\begin{table}
	\caption{\label{tab:EBD212} Elementary Band decomposition for space group $P4_332$ no. 212 and Wyckoff positions $\uw_a = [\frac{\textrm{1}}{\textrm{8}},\frac{\textrm{1}}{\textrm{8}},\frac{\textrm{1}}{\textrm{8}}]$ and $\uw_c = [r, r, r]$. The character tables for the local symmetry groups of the wavevectors listed and $\mathsf{Stab}(\uw_a)$ are reported in Appendix~\ref{sec:AppxCC}.}
	\begin{tabular}{@{}lllll}
	\hline
	 & $\Gamma$ & $X$ & $M$ & $R$ \\
	\hline
	$(\uw_a, a_1)$ & $\Gamma_{a1} \bigoplus \Gamma_{t4}$ & $X_{e5} \bigoplus X_{e7}$ & $M_{a3} \bigoplus M_{a5} \bigoplus M_{e10}$ & $R_{f8}$\\ 
	$(\uw_a, a_2)$ & $\Gamma_{a2} \bigoplus \Gamma_{t5}$ & $X_{e5} \bigoplus X_{e7}$ & $M_{a2} \bigoplus M_{a6} \bigoplus M_{e10}$ & $R_{f8}$\\ 
	$(\uw_a, e_3)$ & $\Gamma_{e3} \bigoplus \Gamma_{t4} \bigoplus \Gamma_{t5}$ & $2 X_{e5} \bigoplus 2 X_{e7}$ & $M_{a2} \bigoplus M_{a3} \bigoplus M_{a5} \bigoplus M_{a6} \bigoplus 2 M_{e10} $ & $R_{e3} \bigoplus R_{e4} \bigoplus R_{f8}$\\
	\hline
	$(\uw_c, a_1)$ & $\Gamma_{a1} \bigoplus \Gamma_{a2} \bigoplus \Gamma_{t4} \bigoplus \Gamma_{t5}$ & $2 X_{e5} \bigoplus 2 X_{e7}$ & $M_{a2} \bigoplus M_{a3} \bigoplus M_{a5} \bigoplus M_{a6} \bigoplus 2 M_{e10}$ & $2 R_{f8}$\\
 $(\uw_c, a_2)$ & $\Gamma_{e3} \bigoplus \Gamma_{t4} \bigoplus \Gamma_{t5}$ & $2 X_{e5} \bigoplus 2 X_{e7}$ & $M_{a2} \bigoplus M_{a3} \bigoplus M_{a5} \bigoplus M_{a6} \bigoplus 2 M_{e10}$ & $R_{e3} \bigoplus R_{e4} \bigoplus R_{f8}$\\
  $(\uw_c, a_3)$ & $\Gamma_{e3} \bigoplus \Gamma_{t4} \bigoplus \Gamma_{t5}$ & $2 X_{e5} \bigoplus 2 X_{e7}$ & $M_{a2} \bigoplus M_{a3} \bigoplus M_{a5} \bigoplus M_{a6} \bigoplus 2 M_{e10}$ & $R_{e3} \bigoplus R_{e4} \bigoplus R_{f8}$\\
 	\hline
	\end{tabular}
	\end{table}

	By considering the EBDs in Tab.~\ref{tab:EBD212} one can make the following observation: along the continuity chord $(\uw_a , e_3)$ the states transforming like $\Gamma_{e3}$ are compatible with the decomposition $2 X_{e5} \bigoplus 2 X_{e7}$, which is also shared by all states along $(\uw_c, a_i)$.
	Thus, at the endpoint $X$ of the dispersion curve, one has that the Bloch states formed either by $d$-GTOs at $\uw_a$ or by $p$-orbitals at $\uw_c$ are compatible.
	Then, the presence of a band inversion along the $\Gamma - X$ path can not be established by checking the symmetry properties of the Bloch states involved, and the typical approach warrants the evaluation of topological invariants at the band crossings, also in light of the presence of two distinct Weyl points (at least for some values of the lattice constant) which could switch the band inversion back to the energy level alignment at the $\Gamma$ point.
	We leave the systematic assessment of topological invariants for future research, since the explicit calculation of the Berry phase, for example, requires the evaluation of higher order derivatives for which a semi-analytical approximation scheme has not yet been developed, however we can rephrase the previous question about the presence of a band inversion in terms of the continuity requirements for the Bloch states involved as a function of the wavevector.
	If one can find two Bloch states that along the dispersion curve can be continuously deformed into one another, then, provided that the corresponding energies are comparable, one could construct a continuous Bloch state interpolating between two different continuity chords, thus leading to a violation of compatibility relations.

\begin{figure}
\includegraphics[width=180mm]{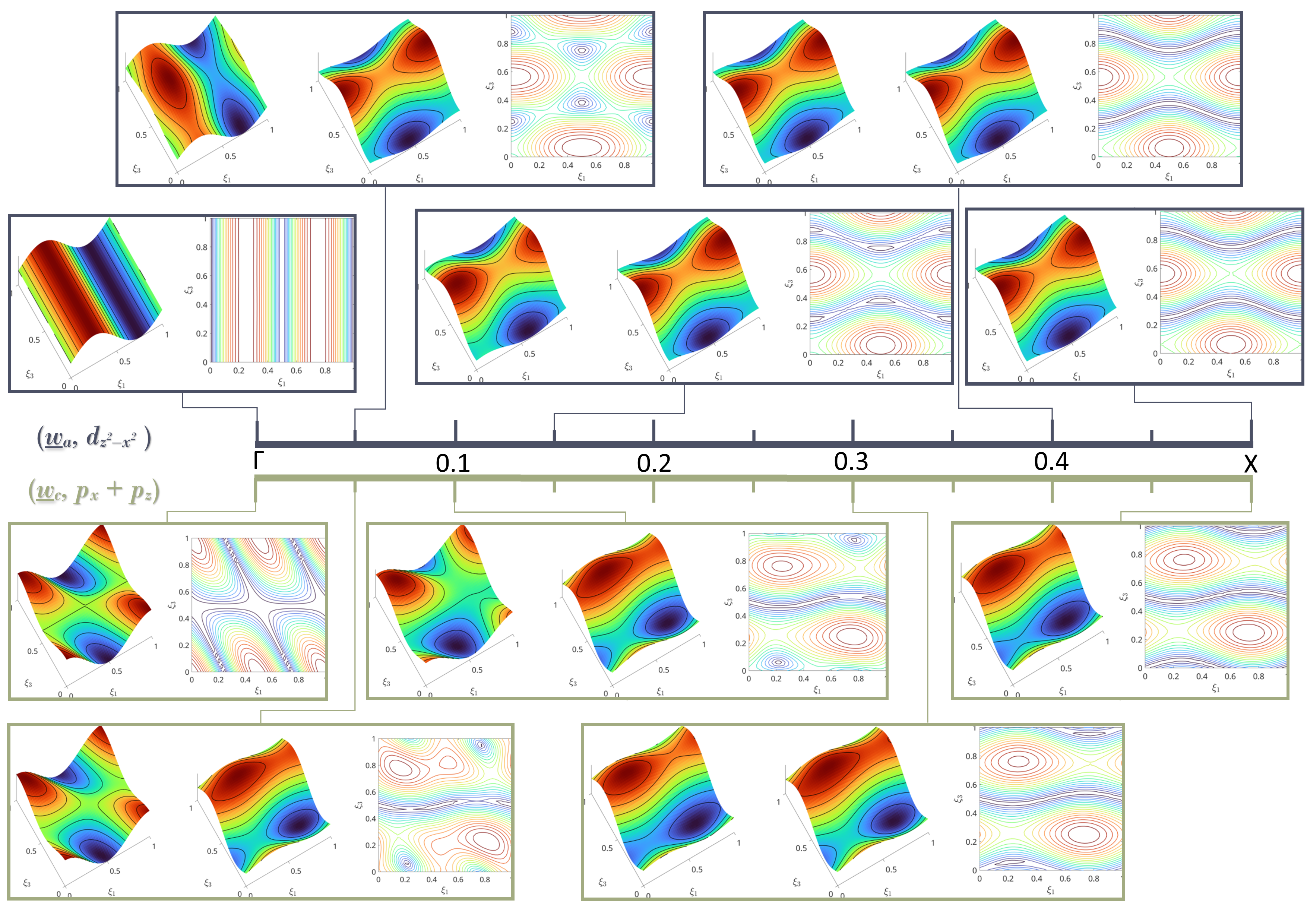}
\caption{\label{fig:BS_dispGX_xi20} Bloch states obtained from the symmetric superposition of $d_{z^2 - x^2}$ GTOs in Wyckoff position $\uw_a$ (blue outlines, top of figure) and from the linear combination $p_x + p_z$ GTOs in Wyckoff position $\uw_c$ (green outlines, bottom of figure) for indicated values of the wavevector along the line $\Gamma - X2$.
Each subplot shows the Bloch state's real and  imaginary (when non-zero) components (as surface plots) and its absolute value (as a contour plot) plotted as a function of the $\uc$ variable in the $\xi_2 = 0$ plane. }
\end{figure}

\begin{figure}
\includegraphics[width=170mm]{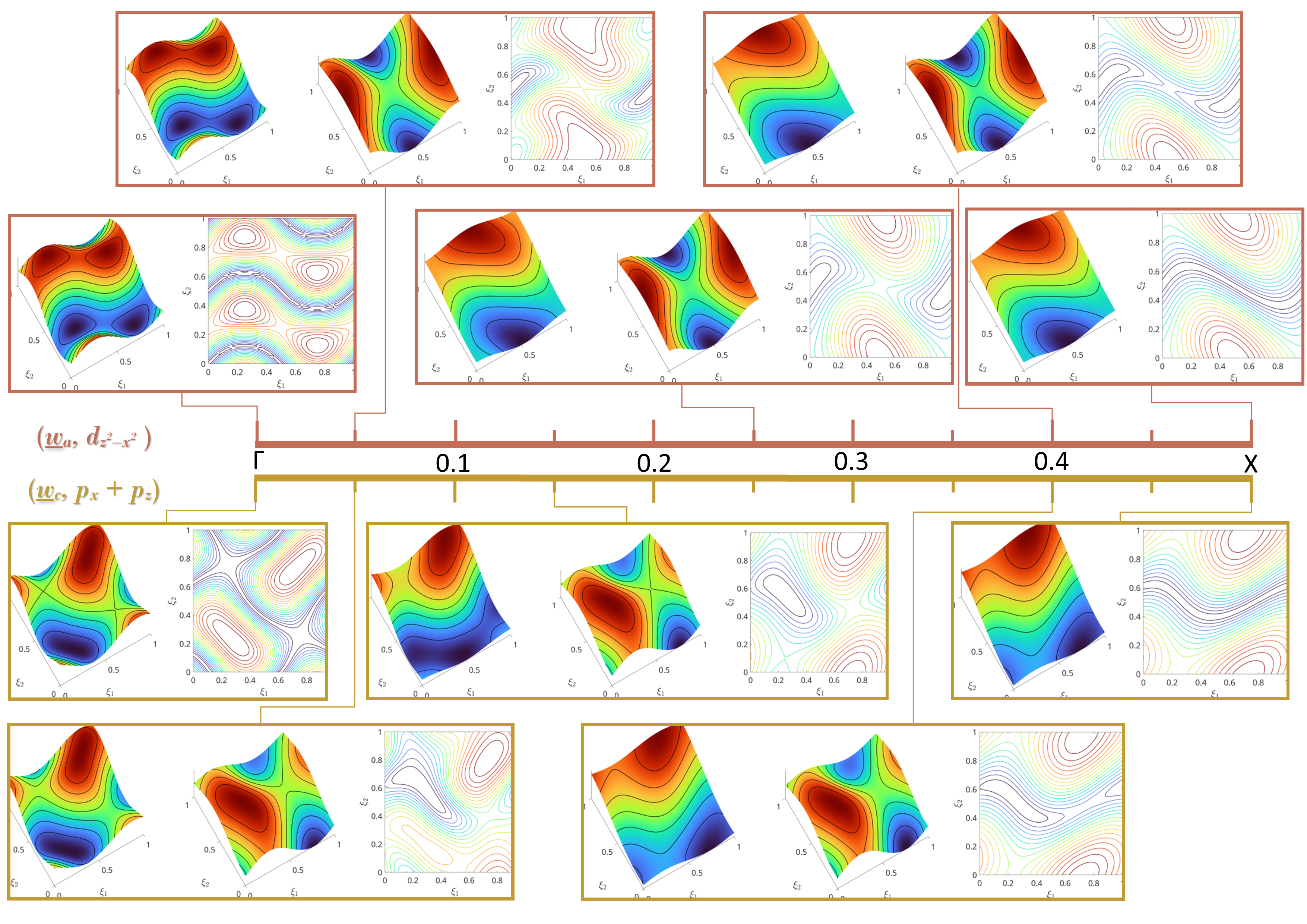}
\caption{\label{fig:BS_dispGX_xi20_planes} Bloch states  as in  the previous picture, with the plots in pink outlines (top of figure) evaluated in the $\xi_3 = 0$ plane and those in yellow outlines (bottom of figure) over the $\xi_3 = \frac{1}{4}$ plane.}
\end{figure}

	To address this question we turn to Fig.~\ref{fig:BS_dispGX_xi20} and Fig.~\ref{fig:BS_dispGX_xi20_planes} where two Bloch states parametrised by different Wyckoff positions (thus pertaining to different continuity chords) are contrasted as the wavevector is varied; the two pictures reproduce different sections of the unit cell.
	Starting with Fig.~\ref{fig:BS_dispGX_xi20}, where we show a section of the unit cell along the $\xi_2 =$0 plane, we can observe  the evolution of a Bloch state associated with the orbitals making up the conduction band (top of figure) \textit{vs} a Bloch state constructed from $p$-orbitals centred on Si atomic positions (bottom of figure) along the $\Gamma - X$ path.
	At the $\Gamma$ point the two states clearly transform like different Irreps, as the wavevector is varied they undergo a substantial rearrangement and for mid values of $\uk$ we can see how the two states become qualitatively similar, with a relative shift by  approximately a vector $\underline{\tau} = [\frac{\textrm{1}}{\textrm{4}}, \textrm{0},  \frac{\textrm{1}}{\textrm{4}}]$, which also impacts the zeroes of the Bloch states.
	Specifically, for the $d$-state one goes from the origin being part of $\Xi_s$ at $\Gamma$ to this no longer being the case for $\uk = X$, and the opposite behaviour is observed for the $p$-state for which the origin $\uc = \uzz$ is part of $\Xi_s$ at the $X$ point.
	The presence of a band inversion would then lead to the violation of compatibility relations, but, at the same time, to the conservation of $\Xi_s$ along the dispersion curve, made possible by the lack of inversion symmetry protecting the one-to-one correspondence established in Appendix~\ref{sec:AppxThm} and we conjecture that this mechanism could be at play in general for the occurrence of a topological band structure.
	To better characterise the two Bloch states considered in this example we plot them along the plane $\xi_3 =\textrm{0}, \frac{\textrm{1}}{\textrm{4}}$ for the $d$- and $p$- state respectively, as shown in Fig.~\ref{fig:BS_dispGX_xi20_planes}.
	We can notice that the two states also have different chirality as the wavevector approaches the $X$ point, we note in passing that the suggested topological phase has been postulated in absence of spin-orbit coupling, hence we assume that states with opposite chirality can be degenerate.

	To identify the interpolating function we proceed as follows: starting from those considered in the figures, we keep the state on the continuity chord $(\uw_c, p_x + p_y)$ fixed and we act with  a translation by a vector $\underline{\tau}$ on $(\uw_a, d_{z^2 - x^2})$.
	We can then relate this action on the coordinate system to that on the Wyckoff position according to Eq.~\ref{eq:Transf_orb}, thus sending $\{ \uw_a \}$ to $\{ \uw_a \} + \underline{\tau}$; the symmetry of the translated Wyckoff position is indeed lower than the original one, with three points belonging to position $\uw_d$ (of which two belong to the same orbit) and one to position $\uw_b$.
	The interpolating function so constructed can then bring the two states considered in this example to overlap, and thus be degenerate.
	We point out that the transformation $(E|\underline{\tau})$ is not an element of the space group in question, nor of its holohedry, as such its action can be assimilated to a symmetry breaking.
	Indeed, when we restore the full symmetry by including not only the translated coordinates $\{\uw_a \} + \underline{\tau}$ in the construction of the interpolating function but also the equivalent points (under the action of the space group operations) in the three disjoint orbits mentioned earlier, the original Bloch state constructed with localised descriptor $(\uw_a, d_{z^2 - x^2})$ is recovered, thus "undoing" the translation $\underline{\tau}$. 
	As such, one is lead to believe that the presence of a band inversion can not be sustained by the states considered without resorting to an interpolating function that breaks additional symmetries in the system, which is offset by the preservation of the set $\Xi_s$ along the energy dispersion curve.
	In turn, this is made possible, as established in Sec.~\ref{sec:TopInvModForms}, by the fact that there is no inversion symmetry that protects the one-to-one correspondence between different pairs $s=(\uw, \uk)$ and the sets $\Xi_s$.

\section{Conclusions}
\label{sec:Concl}

	In the present study a theoretical framework has been proposed that can help systematise the discovery of topological materials for which the single particle picture provides a reasonable approximation to their electronic structure: it can integrate information inferred from the explicit construction of Bloch states from localised orbitals and from the continuity chord label by direct evaluation of the symmetry properties.
	Furthermore, mathematical properties of the analytic complex functions employed to represent the Bloch states are exploited to conjecture the existence of a general mechanism at play when a band inversion takes place.

	Starting from the description of the electronic structure in terms of GTOs we have obtained an expression for the associated Bloch states that are  \textit{entire} functions of the complex variable $\uz = \uk - \OO \uc$.
	The mapping from the description of the material's electronic structure in terms of the single particle orbitals to that in terms of the Bloch states is, in fact, exact and the only approximation resides in the choice of Gaussian orbitals for the expansion of the electronic wavefunction. 
	Details of the material's geometry are taken into account thanks to the period matrix $\OO$ that includes the crystal metric tensor and captures the interplay with the GTO broadening $\beta$, as such the formalism is expected to be able to capture the realistic description of a given material's electronic structure.

	The proposed method is also expected to be robust in regard to the wavevector dependence: the $\vartheta -$functions being \textit{entire} complex functions, in particular, means that the Bloch states can be propagated along the energy dispersion curve in the Brillouin zone and their transformation properties can be efficiently  contrasted with the compatibility relations between Irreps of $G_{\uk}$ for different wavevectors. 
	On the other hand, when composite bands are present, the symmetry analysis alone can not identify the presence of a symmetry enforced band crossing; we have shown this principle for the case of space group no. 207 where compatibility relations alone are not sufficient, since the  continuity chord spans the whole set of energy levels involved, also once the degeneracy among them is lifted.
	In this case the additional labeling provided by the GTO employed in constructing the Bloch state is essential for such identification. 
	
	Besides the computational applicability of the proposed method to real materials, the representation of the Bloch state in terms of $\vartheta -$functions and its derivatives allows us to propose an overarching mechanism, that we conjecture plays a role in the formation of topologically non-trivial band structures.
	We have derived an explicit expression for the parity eigenvalue, making use only of the symmetry properties of the $\vartheta -$functions as modular forms, from this, we have argued that the sets of zeroes of the $\vartheta -$function, $\Xi_s$, could be employed to characterise crystalline topological insulators. 
	This interpretation is exemplified in Sec.~\ref{sec:SG212} where, owing to the lack of inversion symmetry in the crystal, the conservation of $\Xi_s$ along the path in reciprocal space is made possible and would be satisfied in the presence of a non-trivial band connectivity.
	
	In this study we have exploited the proposed scheme to provide an intuitive picture by using a minimal description of the given material; by extension, this approach could allow for a quick scan of the "chemical space" of candidate materials with non-trivial topological properties before resorting to more demanding electronic structure calculations, thus making it a candidate for high-throughput computational studies.
	By the same token, a more refined description of a material's electronic structure can be accommodated straightforwardly  by simply including more GTOs in the construction of the Bloch states, thus making the proposed method suitable to be interfaced with \textit{ab initio} electronic structure codes employing Gaussian basis sets.

\section*{Acknowledgments}
The author wishes to thank Dr. Jan M. Tomczak for the feedback on the manuscript.
	The author acknowledges funding from the Scuola Superiore Meridionale per DM MUR n. 581 24/06/2022. Part of the research presented was carried out with support from the Austrian Science Fund (FWF) through project BandITT P 33571.

\section*{References}
\bibliography{Books.bib, GeneralTheory.bib, GroupTheory.bib, NumericalMethods.bib, Topology.bib, Transport.bib}

\appendix

\section{Properties of $\vartheta -$functions, its derivatives and Bloch states}
\label{sec:Appx}
\begin{defn}
The Riemann $\vartheta -$function is defined as:
\begin{align}
\label{eq:theta}
\vartheta(\uz|\OO) = \sum_{\un \in \Z^g} e^{\ip  \un \cdot \OO \un} e^{2\ip  \un \cdot \uz}
\end{align}
with $\uz \in \C^g$ and $\OO$ is a $g \times g$ symmetric matrix with $\Im \{ \OO\} > 0$, this requirement is necessary for the series on the r. h. s. of Eq.~\ref{eq:theta} to be \textit{absolutely convergent};
the expression $\un \cdot \OO \un$ gets interpreted as the sequential product of the vectors $^t\un  (\OO \un)$.
\end{defn}

	The matrix $\OO$ is an element of the Siegel upper half plane \cite{Mumford1983}
which generalises the complex upper-half plane to the $g$-dimensional case.
$\OO$ is called a period matrix because it can be used to construct the lattice $L_\OO = \Z^g + \OO \Z^g$: in Proposition \ref{prop:quasiperiod} the quasi-periodicity properties of the Riemann $\vartheta -$functions with respect to such lattice are recovered.
For practical purposes $\OO$ encodes some information about the crystal lattice since it is defined as: $\OO = \frac{\ii \beta}{\pi} \mathbf{G}$, where $\mathbf{G}$ is the lattice metric tensor and $\beta$ is the GTOs broadening.

\begin{defn}
The   Riemann $\vartheta -$function with rational characteristics is defined as:
\begin{align}
\label{eq:rattheta}
\ratthe{\ua}{\ub}(\uz|\OO) = e^{\ip  \ua \cdot  \OO \ua} e^{2\ip   \ua \cdot (\uz + \ub)} \vartheta(\uz + \OO \ua  + \ub|\OO).
\end{align}
We will be interested specifically in the case $\ua = \uw$ being the Wyckoff position's coordinates in the $g$-dimensional space, and $\ub = \uzz$.
\end{defn}

\begin{prop}[Quasi-periodicity of the Riemann $\vartheta -$functions]\hfill 
\hfill

In what follows $\um \in \Z^g$.
\label{prop:quasiperiod}
\begin{enumerate}
\item{$\vartheta(\uz + \um | \OO) = \vartheta(\uz | \OO) \; \; \forall \, \um \in \Z^g$.\\ \break
The equality is quite obvious, since by applying the definition one has:
\begin{align*}
\vartheta(\uz + \um | \OO) =\sum_{\un \in \Z^g} e^{\ip  \un \cdot  \OO \un} \, e^{2\ip  \un \cdot (\uz + \um)} = \sum_{\un \in \Z^g} e^{\ip  \un \cdot \OO \un} \, e^{2\ip  \un \cdot \uz} \,  \underbrace{e^{2\ip  \un \cdot \um}}_{=1} = \vartheta(\uz | \OO).
\end{align*}
}

\item{$\vartheta(\uz + \OO\um | \OO) = e^{-\ip  \um \cdot \OO \um } \, e^{-2\ip   \um \cdot \uz} \vartheta(\uz|\OO)$.\\ \break
This relation holds because $\OO$ is symmetric: a representative term in the summation  of the left hand side is:
\begin{align*}
e^{\ip  \un \cdot \OO \un} \, e^{2\ip  \un \cdot (\uz + \OO \um)} = e^{\ip  \un \cdot \OO \un} \, e^{2\ip  \un \cdot \uz} \, e^{\ip   \un \cdot \OO \um} \, e^{\ip  \um \cdot \OO \un} \\
= e^{\ip  (\un + \um) \cdot \OO (\un + \um)} \, e^{2\ip  (\un + \um)\cdot \uz} \, e^{-\ip  \um \cdot \OO \um } \, e^{-2\ip  \um \cdot \uz},
\end{align*}
as $\un$ varies over $\Z^g$ so does $\un + \um$, hence the conclusion follows.}

\item{$\ratthe{\ua+\un}{\ub+\um}(\uz|\OO) = e^{2\ip  \ua \cdot \um} \ratthe{\ua}{\ub}(\uz| \OO)$.\\ \break
Using the Eq.~\ref{eq:rattheta} for the translated characteristics one gets:
\begin{align}
\ratthe{\ua+\un}{\ub+\um}(\uz|\OO) = e^{\ip  (\ua + \un) \cdot \OO (\ua + \un)} \, e^{2\ip  (\ua + \un)\cdot (\uz + \ub +\um)} \, \vartheta(\uz + \OO \ua + \ub + \OO \un  + \um | \OO).
\label{eq:ratthesh}
\end{align}
Using the properties (i) and (ii) above one gets:
\begin{align*}
\vartheta(\uz + \OO \ua + \ub + \OO \un  + \um | \OO) = e^{-\ip  \un \cdot \OO \un} \, e^{-2\ip  \un \cdot (\uz + \OO \ua + \ub)} \, \vartheta(\uz +\OO \ua + \ub|\OO).
\end{align*}
The exponentials in Eq.~\ref{eq:ratthesh} can be expressed as follows:
\begin{align*}
e^{\ip  (\ua + \un) \cdot \OO (\ua + \un)} &= e^{\ip  \ua \cdot \OO \ua } \, e^{\ip  \un \cdot \OO \un } \, e^{2\ip  \un \cdot \OO \ua} \\
e^{2\ip  (\ua + \un)\cdot (\uz + \ub +\um)} &= e^{2\ip  \ua \cdot (\uz +\ub)} \, e^{2\ip  \ua\cdot  \um} \underbrace{e^{2\ip  \un \cdot \um}}_{=1} \, e^{2\ip  \un \cdot \ub} \, e^{2\ip  \un \cdot \uz}.
\end{align*}

Then one can rewrite Eq.~\ref{eq:ratthesh} as:
\begin{align*}
\ratthe{\ua+\un}{\ub+\um}(\uz|\OO) &= e^{2\ip  \ua\cdot \um} \, e^{\ip  \ua \cdot \OO \ua} \, e^{2\ip  \ua \cdot (\uz +\ub)} \vartheta(\uz + \OO \ua + \ub |\OO)\\
& = e^{2\ip  \ua \cdot \um }\, \ratthe{\ua}{\ub}(\uz|\OO).
\end{align*}
}

\item{$\ratthe{\ua}{\ub}(\uz + \um | \OO)= e^{2\ip  \ua \cdot \um} \ratthe{\ua}{\ub}(\uz| \OO)$. \\ \break
Applying the definition one gets to the result more laboriously than by just noticing that 
\begin{align*}
\ratthe{\ua}{\ub}(\uz + \um | \OO)=\ratthe{\ua}{\ub + \um}(\uz |\OO)
\end{align*}
and then using property 3. above. }

\item{$\ratthe{\ua}{\ub}(\uz + \OO \um |\OO) = e^{-\ip  \um \cdot \OO \um} \, e^{-2\ip  \um \cdot (\uz +\ub)} \, \ratthe{\ua}{\ub}(\uz|\OO)$.\\ \break
In this case one uses the definition in Eq.~\ref{eq:rattheta} for the Riemann $\vartheta -$function with rational characteristics and the fact that $\OO$ is a symmetric matrix:
\begin{align*}
\ratthe{\ua}{\ub}(\uz + \OO \um |\OO) &
= e^{\ip  \ua \cdot \OO \ua} \, e^{2\ip  \ua\cdot (\uz +\OO \um +\ub)} \, \vartheta(\uz+ \OO \ua +\OO \um +\ub |\OO) \\
& = e^{\ip  \ua \cdot \OO \ua} \, e^{2\ip  \ua\cdot (\uz +\ub)} \, e^{2\ip  \ua \cdot \OO \um} \, e^{-\ip  \um \cdot \OO \um} \, e^{-2\ip  \um\cdot (\uz + \OO \ua + \ub)} \, \vartheta(\uz + \OO \ua +\ub|\OO)\\
& = e^{2\ip  \ua \cdot \OO \um} \,e^{-\ip  \um \cdot \OO \um} \,e^{-2\ip  \um\cdot (\uz + \OO \ua + \ub)} \, \ratthe{\ua}{\ub}(\uz|\OO)\\
& =e^{-\ip  \um \cdot \OO \um}\,e^{-2\ip  \um\cdot (\uz + \ub)} \, \cancel{e^{2\ip  \ua \cdot \OO \um}} \, \cancel{e^{-2\ip  \ua \cdot \OO \um}} \, \ratthe{\ua}{\ub}(\uz|\OO),
\end{align*}
where in going to the second line we have used property 2. above and for the last line the fact that $\OO =  {^t\OO}$ has been exploited, which leaves the inner product invariant.
}
\end{enumerate}
\end{prop}

\begin{prop}[Action of space group operations]
If $F= (P|\uu)$ is the space group operation, with $P \in \Z^{g \times g}$ a rotation matrix and $\uu \in \Q^g$ the non-symmorphic translation vector\footnote{ thanks to the Affine embedding theorem it is always possible to express the non-symmorphic translation as a rational vector \cite{Eick2006}}, one has that in order to define an action of the space group operations on the functions over $\R^g$ one must have $F[f(\ux)] = f(F^{-1} \ux)$.
Thus, for the action on the Riemann $\vartheta -$functions we have:
\begin{align}
F \left [ \ratthe{\uw}{\uzz}(\uk - \OO \uc| \OO) \right ] = e^{- \ii \pi   \uu \cdot \tO \uu}\, e^{- 2 \pi \ii  \uu \cdot (^t P^{-1}\uk - \tO \uc)} \, \ratthe{\uv + \uu}{\uzz}(^tP^{-1}\uk - \tO \uc| \tO), \label{eq:rattheTransf}
\end{align}
that is: the Riemann $\vartheta -$function gets centred at the new Wyckoff position $\uv + \uu = P\uw + \uu$, while the matrix $\OO$ gets transformed by similarity into $\tO = {^t P}^{-1} \OO P^{-1}$.
\begin{proof}
The terms to consider are given in Eq.~\ref{eq:rattheta} and they transform like:
\begin{enumerate}[1.]
\item{$ \uw \cdot \OO \uw \longrightarrow { (P^{-1} P\uw) \cdot \OO (P^{-1} P\uw ) =  {^t(P\uw)}\, {^tP^{-1}}\OO P^{-1} (P\uw) = \uv \cdot \tO \uv } $}.
\item{$ \uw \cdot (\uk - \OO \uc) \longrightarrow { ^t(}P^{-1}\uw ) ( {^tP^{-1}} \uk - {^tP^{-1}}\OO P^{-1} ( \uc - \uu)) = \uv  \cdot ({^tP^{-1}}\uk - \tO \uc + \tO \uu)$}. 
\item{${ \un} \cdot \OO \un\ \longrightarrow {^t(P^{-1} P  \un)} \, \OO  (P^{-1} P \un) = {^t(P\un)} \, {^tP^{-1}} \OO P^{-1} (P\un) = \um \cdot \tO \um  $}.
\item{${ \un} \cdot (\uk - \OO \uc + \OO \uw) \longrightarrow { ^t(P\un)} \cdot ({^tP^{-1}\uk} - \tO \uc + \tO \uu + \tO (P\uw)) = \um \cdot ({^t P^{-1}\uk} \, - \tO \uc + \tO (\uv + \uu)) $}.
\end{enumerate}
	Note that the vectors $\un $ and $\um $ span the same lattice, since the rotations sending one to the other belong to a subgroup of the holohedry, hence, by replacing the terms above in the definition of the Riemann $\vartheta -$function with rational characteristics and summing over the lattice translations $\um$ we get:
\begin{align}
\textrm{l. h. s. Eq. (\ref{eq:rattheTransf})} = & e^{\ii \pi { \uv} \cdot \tO \uv} \, e ^{2 \pi \ii  \uv \cdot ({^tP^{-1}\uk} - \tO \uc + \tO \uu)} \,
\sum_{\um \in \Z^g} e^{\ii \pi  \um \cdot \tO \um}\, e^{2 \pi \ii  \um \cdot ({^tP^{-1}\uk} - \tO \uc + \tO (\uv + \uu))} 
\nonumber \\
= & e^{\ii \pi   \uv \cdot \tO \uv}\, e ^{2 \pi \ii   \uv \cdot \tO \uu} \, e^{2 \pi \ii  \uv \cdot ({^tP^{-1}\uk} - \tO \uc)}\,  \vartheta({^tP^{-1}\uk} - \tO \uc + \tO(\uv + \uu)|\tO) \nonumber \\
= & e^{\ii \pi  (\uv + \uu) \cdot \tO (\uv + \uu)} \, e^{- \ii \pi   \uu \cdot \tO \uu} \, e^{2 \pi \ii  (\uv + \uu) \cdot ({^tP^{-1}\uk} - \tO \uc)} \, e^{-2 \pi \ii   \uu \cdot ({^tP^{-1}\uk} - \tO \uc)}\, \vartheta({^tP^{-1}\uk} - \tO \uc + \tO(\uv + \uu)|\tO) \nonumber \\
= & e^{-\ii \pi  \uu \cdot \tO \uu}\, e^{-2 \pi \ii   \uu \cdot ({^tP^{-1}\uk} - \tO \uc)}\, \ratthe{\uv + \uu}{\uzz}({^tP^{-1}\uk} - \tO \uc|\tO)
\end{align}
\end{proof}
\end{prop}

\subsection{First derivative of $\vartheta -$functions with rational characteristics}

\begin{defn}
\label{defn:D1theta}
The directional derivative along the direction $\langle \ue |$ with respect to the complex variable $\uz$ of a   Riemann $\vartheta -$function is defined as: 
\begin{align}
D_{\ed} (\uz) \vartheta(\uz + \OO \ua | \OO) & = D_{\ed}(\uz) \sum_{\un \in \Z^g} e^{\ii \pi   \un \cdot \OO \un} \, e^{2\pi \ii   \un \cdot (\uz + \OO \ua)} \nonumber \\
& = 2\pi \ii \sum_{\un \in \Z^g}  \langle \ue | \un \rangle e^{\ii \pi   \un \cdot \OO \un} \, e^{2\pi \ii   \un \cdot (\uz + \OO \ua)},
\end{align}
\end{defn}

The definition above can be used in the ensuing evaluation of the derivative of the Riemann $\vartheta -$function with rational characteristics:
\begin{align}
D_{\ed} (\uk - \OO \uc) \ratthe{\uw}{\uzz}(\uk - \OO \uc | \OO)= & 2\pi \ii \ed \uw \rangle \, \ratthe{\uw}{\uzz}(\uk - \OO \uc | \OO) + e^{\ii \pi   \uw \cdot \OO \uw} e^{2\pi \ii   \uw \cdot (\uk - \OO \uc)} \nonumber \\
& \times  D_{\ed} (\uk - \OO \uc) \vartheta (\uk - \OO \uc + \OO \uw | \OO).
\label{eq:D1ratthe}
\end{align}

\begin{prop}[Action of space group operations]
For a space group operation $F=(P|\uu)$ with $P$ being a rotation and $\uu$ the non-symmorphic translation (possibly present) the action on the derivative of a   Riemann $\vartheta -$function with rational characteristics evaluates as:
\begin{align}
F  \left [  D_{\ed} (\uk - \OO \uc) \ratthe{\uw}{\uzz}(\uk - \OO \uc | \OO) \right] = & e^{- \ii \pi   \uu \cdot \tO \uu } \, e^{-2\pi \ii   \uu \cdot \utz} \left \{
D_{\langle \ue P^{-1}|} (\utz) \ratthe{\uv + \uu}{\uzz}(\utz | \tO) \right. \nonumber \\
& \left . - 2\pi \ii \, \langle \ue P^{-1}|  \uu \rangle \, \ratthe{\uv + \uu }{\uzz} (\utz|\tO) \right \},
\end{align}
with $\tO = {^tP^{-1}} \OO P^{-1}$ the transformed counterpart of the symmetric complex matrix $\OO$ and $\utz = {^tP^{-1}\uk - \tO \uc}$ the transformed complex variable.
\begin{proof}
First notice that thanks to Eq.~\ref{eq:D1ratthe} we can separate out two terms:
\begin{align}
F  \left [  D_{\ed} (\uk - \OO \uc) \ratthe{\uw}{\uzz}(\uk - \OO \uc | \OO) \right] = & F  \left [ 2 \pi \ii \ed \uw \rangle \ratthe{\uw}{\uzz}(\uk - \OO \uc | \OO) + \right . \nonumber \\
 & \left . + e^{\ii \pi \uw \cdot \OO \uw}  \, e^{2\pi \ii \uw \cdot (\uk - \OO \uc)} D_{\ed} (\uk - \OO \uc) \vartheta(\uk - \OO \uc + \OO \uw| \OO) \right] 
\end{align}
and then study each of them separately under the action of the space group operations.
\begin{enumerate}[1.]
\item{$F \left [ 2 \pi \ii \ed \uw \rangle \ratthe{\uw}{\uzz}(\uk - \OO \uc | \OO) \right ]$.\\
For the inner product we have the following transformation rule:
\begin{align}
\ed \uw \rangle \longrightarrow \langle \ue | P^{-1} P  \uw \rangle = \langle \ue P^{-1} | \uv \rangle
\end{align}
whereas the Riemann $\vartheta -$function with rational characteristics transforms as reported in Eq.~\ref{eq:rattheTransf},
hence we have the contribution below:
\begin{align}
2 \pi \ii \langle \ue P^{-1} | \uv \rangle e^{-\ii \pi \uu \cdot \tO \uu } \, e^{-2\pi \ii   \uu \cdot ({^tP^{-1}\uk} - \tO \uc)} \ratthe{\uv + \uu}{\uzz}({^tP^{-1}\uk} - \tO \uc | \tO). 
\end{align}
}
\item{$F \left [ e^{\ii \pi \uw \cdot \OO \uw}  \, e^{2\pi \ii \uw \cdot (\uk - \OO \uc)} D_{\ed} (\uk - \OO \uc) \vartheta(\uk - \OO \uc + \OO \uw| \OO) \right ]$.\\
By applying Definition \ref{defn:D1theta} the transform of the directional derivative reads as: 
\begin{align}
F \left [ 2 \pi \ii \sum_{\un \in \Z^g} \ed \un \rangle e^{\ii \pi \un \cdot \OO \un } \, e^{2\pi \ii  \un \cdot (\uk - \OO \uc + \OO \uw)} \right ] &= 
2\pi \ii  \sum_{\um \in \Z^g} \langle \ue P^{-1}| \um \rangle \, e^{\ii \pi  \um \cdot \tO \um} \, e^{2\pi \ii  \um \cdot ({^tP^{-1}\uk} - \tO \uc + \tO \uv + \tO \uu)} \nonumber \\
& = D_{\langle \ue P^{-1}|} (\utz) [\vartheta(\utz + \tO(\uv+\uu)|\tO)]
\end{align}
with $\um = P \un , \, \uv = P\uw$ and the prefactor transforms as follows:
\begin{align}
F \left [ e^{\ii \pi \uw \cdot \OO \uw} \, e^{2\pi \ii \uw \cdot (\uk - \OO \uc)}\right] & = 
e^{\ii \pi  \uv \cdot \tO \uv} \, e^{2\pi \ii  (\uv) \cdot ({^tP^{-1}}\uk - \tO \uc + \tO \uu)}.
\end{align}
}
\end{enumerate}

Bringing the contributions from the transformed terms 1. and 2. above we get:
\begin{align}
& F  \left [  D_{\ed} (\uk - \OO \uc) \ratthe{\uw}{\uzz}(\uk - \OO \uc | \OO) \right] = \nonumber \\
= & 2 \pi \ii \, \langle \ue P^{-1} | \uv \rangle \, e^{- \ii \pi \uu \cdot \tO \uu } \, e^{-2 \pi \ii \uu \cdot ({^tP^{-1}}\uk - \tO \uc)} \, \ratthe{\uv + \uu}{\uzz} ({^tP^{-1}}\uk - \tO \uc | \tO) + \nonumber \\
+ & e^{\ii \pi \uv \cdot \tO \uv } \, e^{2 \pi \ii \uv \cdot \tO \uu } \, e^{2\pi \ii \uv \cdot ({^tP^{-1}}\uk - \tO \uc)} D_{\langle \ue P^{-1}} \vartheta({^tP^{-1}}\uk - \tO \uc +\tO (\uv + \uu) |\tO). \label{eq:D1transf1}
\end{align}
In the expression above one can identify the transform of a directional derivative as follows:
\begin{align}
\textrm{Eq. (\ref{eq:D1transf1})} &= e^{- \ii \pi \uu \cdot \tO \uu } \, e^{-2 \pi \ii \uu \cdot ({^tP^{-1}}\uk - \tO \uc)}  \left \{ 2 \pi \ii \langle \ue P^{-1} | \uv \rangle \, \ratthe{\uv + \uu}{\uzz} ({^tP^{-1}}\uk - \tO \uc | \tO) \right . \nonumber \\
& \left. + e^{\ii \pi (\uv + \uu ) \cdot \tO (\uv + \uu )} \, e^{2\pi \ii  (\uv + \uu) \cdot ({^tP^{-1}}\uk - \tO \uc)} \, D_{\langle \ue P^{-1}|} ({^tP^{-1}\uk} - \tO \uc) \vartheta \left ( {^tP^{-1}\uk} - \tO \uc + \tO (\uv + \uu) | \tO \right )   \right \} \nonumber \\
& = e^{- \ii \pi   \uu \cdot \tO \uu } \, e^{-2\pi \ii   \uu \cdot ({^tP^{-1}\uk} - \tO \uc)} \left \{
D_{\langle \ue P^{-1}|} ({^tP^{-1}\uk} - \tO \uc) \ratthe{\uv + \uu}{\uzz}({^tP^{-1}\uk} - \tO \uc | \tO) \right. \nonumber \\
& \left . - 2\pi \ii \langle \ue P^{-1}|  \uu \rangle \ratthe{\uv + \uu }{\uzz} ({^tP^{-1}\uk} - \tO \uc) \right \}.
\label{eq:D1transf}
\end{align}
\end{proof}

Notice that the derivative of a   Riemann $\vartheta -$function with rational characteristics transforms into the transformed derivative only if the translation $\uu$ is orthogonal to the direction $ \langle \ue P^{-1}| $ along which the derivative is evaluated, otherwise an additional contribution proportional to the Riemann $\vartheta -$function weighted by the projection $\langle \ue P^{-1} | \uu \rangle$ crops up.

\end{prop}

\subsection{Second derivative of $\vartheta -$functions with rational characteristics}
Let us start with the following:
\begin{defn} \label{defn:2ndDer}
For the   Riemann $\vartheta -$function one has: 
\begin{align}
\DD (\uz) [\vartheta(\uz| \OO)]= (2 \pi \ii)^2 \sum_{\un \in \Z^g} \langle \ue_1 |\un \rangle \langle \ue_2 | \un \rangle e^{\ii \pi \un \cdot \OO \un } \, e^{2\pi \ii \un \cdot \uz} 
\end{align}
with the obvious property $\DD = D^{(2)}_{\langle \ue_2| \, \langle \ue_1|}$.
We are going to omit the brackets when the function being derived is obvious from the context and the variable of derivation further down in the manuscript.
\end{defn}

The expression for the second derivative of the Riemann $\vartheta -$function with rational characteristics then can be worked out as follows:
\begin{align}
\DD (\uk - \OO \uc) \ratthe{\uw}{\uzz}(\uk - \OO \uc|\OO) = \DD (\uk - \OO \uc) \left [ e^{\ii \pi  \uw \cdot \OO \uw} \, e^{2\pi \ii \uw \cdot (\uk - \OO \uc)} \vartheta(\uk - \OO \uc + \OO \uw |\OO) \right ] \nonumber \\
= e^{\ii \pi  \uw \cdot \OO \uw} \, \left \{ \DD \left [ e^{2\ii\pi  \uw \cdot (\uk - \OO \uc)}\right ] \vartheta(\uk - \OO \uc + \OO \uw |\OO) 
+ 
D_{\langle \ue_1 |} \left [ e^{2\pi \ii  \uw \cdot (\uk - \OO \uc)}\right ] D_{\langle \ue_2 |} \left [ \vartheta(\uk - \OO \uc + \OO \uw|\OO) \right ] \right . \nonumber \\
+
\left . D_{\langle \ue_2 |} \left [ e^{2\pi \ii  \uw \cdot (\uk - \OO \uc)}\right ] D_{\langle \ue_1 |} \left [ \vartheta(\uk - \OO \uc + \OO \uw|\OO) \right ]
+
e^{2\pi \ii  \uw \cdot (\uk - \OO \uc)} \DD \left [ \vartheta (\uk - \OO \uc + \OO \uw | \OO) \right ]
\right \}. \label{eq:D2rattheDef}
\end{align}

\begin{prop}[Action of space group operations]
For a space group operation defined as in the previous Proposition, we want to establish the transformation rule for the second derivative of a Riemann $\vartheta -$function with rational characteristics.
The final formula reads:
\begin{align}
F \left [ \DD (\uk - \OO \uc) \ratthe{\uw}{\uzz}(\uk - \OO \uc|\OO)\right] = &
e^{-\ii \pi  \uu \cdot \tO \uu} \DDt (\uk - \tO \uc) \left [ e^{-2\pi \ii  \uu \cdot (\uk - \tO \uc)} \right . \nonumber \\
& \left . \ratthe{\uw - \uu}{\uzz} (\uk - \tO \uc| \tO)\right ]. \label{eq:D2transf}
\end{align}
\begin{proof}
In Eq.~\ref{eq:D2rattheDef} there are four terms that can be idenfied and studied independently of the prefactor:
\begin{align}
e^{\ii \pi  \uw \cdot \OO \uw} \longrightarrow e^{\ii \pi  \uv \cdot \tO \uv} 
\label{eq:prefact}
\end{align}
\begin{enumerate}[1.]

\item{$\DD \left [ e^{2\ii\pi \uw \cdot (\uk - \OO \uc)}\right ] \vartheta(\uk - \OO \uc + \OO \uw |\OO)$.\\
The first term can be rewritten immediately as follows:
\begin{align}
(2\pi \ii)^2 \langle \ue_1 | \uw \rangle \langle \ue_2 | \uw \rangle e^{2\pi \ii  \uw \cdot (\uk - \OO \uc)} \vartheta(\uk - \OO \uc + \OO \uw| \OO)
\end{align}
and then it transforms into:
\begin{align}
(2\pi \ii )^2 \, \langle \ue_1 P^{-1} | \uv\rangle \langle \ue_2 P^{-1} | \uv\rangle 
e^{2\pi \ii \uv \cdot ({^tP^{-1}\uk} - \tO \uc + \tO \uc)} \, F[\vartheta(\uk - \OO\uc + \OO \uw|\OO)]
\end{align}
where the transform of the Riemann $\vartheta -$function reads as:
\begin{align}
F[\vartheta(\uk - \OO\uc + \OO \uw|\OO)] = \sum_{\um \in \Z^g} e^{\ii \pi \um \cdot \tO \um} \, e^{-2\pi \ii \um \cdot ({^tP^{-1}}\uk - \tO \uc + \tO(\uv+\uu)} = \vartheta({^tP^{-1}\uk} -\tO \uc + \tO(\uv + \uu)|\tO).
\end{align}
By multiplying in the prefactor in Eq.~\ref{eq:prefact} this term can be expressed as:
\begin{align}
(2\pi \ii )^2  \langle \ue_1 P^{-1} | \uv\rangle \langle \ue_2 P^{-1} | \uv \rangle \,
e^{\ii \pi \uv \cdot \tO \uv} \, e^{2\pi \ii  \uv \cdot \tO \uu} \, e^{2 \pi \ii \uv \cdot \utz} \, \vartheta(\utz + \tO(\uv + \uu)|\tO) \nonumber \\
= (2\pi \ii )^2  \langle \ue_1 P^{-1} | \uv\rangle \langle \ue_2 P^{-1} | \uv \rangle \,
e^{- \ii \pi \uu \cdot \tO \uu } \, e^{\ii \pi (\uv + \uu) \cdot \tO (\uv +\uu)} \, e^{-2 \pi \ii \uu \cdot \utz} \, e^{2 \pi \ii (\uv + \uu) \cdot \utz} \, \vartheta(\utz + \tO(\uv + \uu)|\tO) \nonumber \\
= (2 \pi \ii )^2 \langle \ue_1 P^{-1} | \uv\rangle \langle \ue_2 P^{-1} | \uv \rangle \, e^{- \ii \pi \uu \cdot \tO \uu } \, e^{-2 \pi \ii \uu \cdot \utz} \, \ratthe{\uv + \uu }{\uzz}(\utz| \tO).
\end{align}
}

\item{$D_{\langle \ue_1 |} \left [ e^{2\pi \ii  \uw \cdot (\uk - \OO \uc)}\right ] D_{\langle \ue_2 |} \left [ \vartheta(\uk - \OO \uc + \OO \uw|\OO) \right ] $.\\
The second term can be written more explicitly as:
\begin{align}
2\pi \ii \langle \ue_1 | \uw \rangle e^{2\pi \ii  \uw \cdot (\uk - \OO \uc)} \times 2\pi \ii \sum_{\un \in \Z^g} \langle \ue_2 | \un \rangle e^{\ii \pi \un \cdot \OO \un} e^{2\pi \ii \un \cdot (\uk - \OO \uc + \OO \uw)}
\end{align}
and the action of the space group operation returns:
\begin{align}
(2\pi \ii)^2  \langle \ue_1 P^{-1}|\uv \rangle  e^{2\pi \ii  \uv \cdot ({^tP^{-1}}\uk - \tO \uc + \tO \uu)} \, \sum_{\um \in \Z^g} \langle \ue_2 P^{-1}|\um \rangle e^{\ii \pi  \um \cdot \tO \um } \, e^{2\pi \ii \um \cdot ({^tP^{-1}\uk} - \tO \uc + \tO( \uv + \uu))}. \label{eq:2ndTerm}
\end{align}
So basically there are two terms: one involving a derivative of the Riemann $\vartheta -$function and the other one being linear in the Riemann $\vartheta -$function:
\begin{align}
\textrm{Eq. (\ref{eq:2ndTerm})}= 
	e^{2 \pi \ii \uv \cdot \tO \uu}  D_{\langle \ue_1P^{-1}|}({^tP^{-1}\uk} - \tO \uc) \left [ e^{2\pi \ii \uv \cdot ({^tP^{-1}\uk} - \tO \uc)} \right ]  D_{\langle \ue_2 P^{-1}|}({^tP^{-1}\uk} - \tO \uc) \left [ \vartheta({^tP^{-1}\uk} - \tO \uc + \tO (\uv +\uu) | \tO) \right].
\end{align}
}

\item{$D_{\langle \ue_2 |} \left [ e^{2\pi \ii  \uw \cdot (\uk - \OO \uc)}\right ] D_{\langle \ue_1 |} \left [ \vartheta(\uk - \OO \uc + \OO \uw|\OO) \right ]$. \\
For the third term simply make the substitution $ 1 \longleftrightarrow 2$ in the previous result.
}
	
\item{$e^{2\pi \ii  \uw \cdot (\uk - \OO \uc)} \DD \left [ \vartheta (\uk - \OO \uc + \OO \uw | \OO) \right ]$.\\
which can be written explicitly by using Definition \ref{defn:2ndDer} as:
\begin{align}
(2\pi \ii )^2 \, e^{2 \pi \ii \uw \cdot (\uk - \OO \uc)} \sum_{\un \in \Z^g} \langle \ue_1| \un \rangle \langle \ue_2| \un \rangle \, e^{\ii \pi  \un \cdot \OO \un} \, e^{2\pi \ii \un \cdot (\uk -\OO \uc + \OO \uw)}.
\end{align}

The action of the space group operation also in this case can be evaluated rather straightforwardly as:
\begin{align}
(2 \pi \ii )^2 \, e^{2 \pi \ii \uv \cdot ({^tP^{-1}} -\tO \uc + \tO \uu)} \, \sum_{\um \in \Z^g}{\langle \ue_1 P^{-1} |\um \rangle \langle \ue_2 P^{-1}| \um \rangle} e^{\ii \pi \um \cdot \tO \um} \, e^{2 \pi \ii \um \cdot ({^t P^{-1} \uk } - \tO \uc + \tO (\uv + \uu))} \nonumber \\
= e^{2 \pi \ii \uv \cdot (\utz + \tO \uu)} \DDt (\utz) \left[ \vartheta(\utz + \tO (\uv + \uu) | \tO) \right].
\end{align}
}
\end{enumerate}

Now we can try to express the action of space group operations on the second derivative of a Riemann $\vartheta -$function with rational characteristics:
\begin{align}
 F \left [ \DD (\uk - \OO \uc) \ratthe{\uw}{\uzz}(\uk - \OO \uc|\OO)\right] & = \nonumber \\
= & e^{- \ii \pi \uv \cdot \tO \uv} \left \{ (2 \pi \ii)^2 \langle\ue_1 P^{-1} | \uv \rangle \langle \ue_2 P^{-1} | \uv \rangle \, e^{2 \pi \ii \uv \cdot (\utz + \tO \uu)} \, \vartheta(\utz + \tO (\uv + \uu)|\tO) \right . \nonumber \\
+ & e^{2 \pi \ii \uv \cdot \tO \uu} D_{\langle \ue_1 P^{-1}|}(\utz) \left [ e^{2 \pi \ii \uv \cdot \utz}\right ]
 D_{\langle \ue_2 P^{-1}|}(\utz) \left [ \vartheta(\utz + \tO (\uv + \uu) | \tO) \right ] \nonumber \\
+ & e^{2 \pi \ii \uv \cdot \tO \uu} D_{\langle \ue_2 P^{-1}|}(\utz) \left [ e^{2 \pi \ii \uv \cdot \utz}\right ]
 D_{\langle \ue_1 P^{-1}|}(\utz) \left [ \vartheta(\utz + \tO (\uv + \uu) | \tO) \right ] \nonumber \\
+ & \left . e^{2 \pi \ii  \uv \cdot (\utz + \tO \uu)} \DDt (\utz) \left [\vartheta(\utz + \tO (\uv + \uu)| \tO) \right] \right \} = \nonumber \\
= & e^{- \ii \pi \uv \cdot \tO \uv} e^{2 \pi \ii \uv \cdot \tO \uu} \left \{ \DDt \left[ e^{2 \pi \ii \uv \cdot \utz }\right] \, \vartheta(\utz + \tO (\uv + \uu)|\tO) \right . \nonumber \\
+ & D_{\langle \ue_1 P^{-1}|}(\utz) \left [ e^{2 \pi \ii \uv \cdot \utz}\right ]
 D_{\langle \ue_2 P^{-1}|}(\utz) \left [ \vartheta(\utz + \tO (\uv + \uu) | \tO) \right ] \nonumber \\
+ & D_{\langle \ue_2 P^{-1}|}(\utz) \left [ e^{2 \pi \ii \uv \cdot \utz}\right ]
 D_{\langle \ue_1 P^{-1}|}(\utz) \left [ \vartheta(\utz + \tO (\uv + \uu) | \tO) \right ] \nonumber \\
+ & \left . e^{2 \pi \ii  \uv \cdot \utz } \DDt (\utz) \left [\vartheta(\utz + \tO (\uv + \uu)| \tO) \right] \right \} = \nonumber \\
= & e^{- \ii \pi \uu \cdot \tO \uu} e^{\ii \pi (\uv +\uu) \tO (\uv + \uu)} \DDt (\utz) \left [ e^{2 \pi \ii \uv  \cdot \utz} \vartheta(\utz + \tO (\uv + \uu)| \tO)\right ] \nonumber \\
= & e^{- \ii \pi \uu \cdot \tO \uu} \DDt (\utz) \left [ e^{- 2 \pi \ii \uu \cdot \utz} \ratthe{\uv + \uu}{\uzz}(\utz | \tO)\right ].
\end{align}
\end{proof}
\end{prop}

\subsection{Transformation properties of Bloch states under space group operations}
\begin{prop}[Transformation properties for $s$-orbital Bloch states]
Here we detail the proof for the derivation of Eq.~\ref{eq:Transf_orb} for the case $I=s$.
\begin{proof}
The components whose transformations are required are those of the theta function with rational characteristics (provided already in Eq.~\ref{eq:rattheTransf}), as well as the following:
\begin{align}
e^{\ii \pi \uc \cdot \OO \uc} &\longrightarrow e^{\ii \pi  (\uc - \uu) \cdot \tO (\uc - \uu)} = e^{\ii \pi  \uc \cdot \tO \uc} \, e^{\ii \pi  \uu \cdot \tO \uu} \, e^{- 2 \pi \ii  \uu \cdot \tO \uc} \\
e^{- 2 \pi \ii \uw \cdot \uk} &\longrightarrow e^{- 2 \pi \ii  {^t(P}\uw ) \cdot {^tP^{-1}}\uk} = e^{- 2 \pi \ii  \uv \cdot {^tP^{-1}}\uk}.
\end{align}
Hence:
\begin{align}
F \left [ \phi_{\langle \uk|}^{(\beta, s)} ( \uc|\uw, \OO)\right] =& N_{\uk}^{(\beta,s)} C^{(\beta,s)} \, e^{\ii \pi \uc \cdot \tO \uc} \, \cancel{e^{\ii \pi \uu \cdot \tO \uu}} \, \bcancel{e^{- 2 \pi \ii \uu \cdot \tO \uc}} \, e^{- 2 \pi \ii \uv \cdot {^tP^{-1}}\uk} \, \cancel{e^{- \ii \pi   \uu \cdot \tO \uu}} \, e^{- 2 \pi \ii  \uu \cdot ({^tP}^{-1}\uk \bcancel{- \tO \uc})} \ratthe{\uv + \uu}{\uzz}({^tP^{-1}}\uk - \tO \uc| \tO) \nonumber \\
& = N_{\uk}^{(\beta,s)} C^{(\beta,s)} \, e^{\ii \pi \uc \cdot \tO \uc} \, e^{- 2 \pi \ii (\uv + \uu) \cdot {^tP}^{-1} \uk} \,\ratthe{\uv + \uu}{\uzz}(^tP^{-1}\uk - \tO \uc| \tO) \\
& = \phi_{\langle \uk P^{-1}|}^{(\beta, s)}( \uc |\uv + \uu, \tO).
\end{align}
\end{proof}
\end{prop}

\begin{prop}[Transformation properties for $p$-orbital Bloch states]
For the $p$-type state,  there are three ingredients required in order to characterise the transformation properties, namely: the form factor $f_i$, the $\phi$ function whose transformation rule has already been given in the proposition above and its derivative.
\begin{align}
f_{\langle \ue |} (\uc) = \langle \ue | \uc \rangle &\longrightarrow \langle \ue P^{-1} | \uc \rangle - \langle \ue P^{-1} | \uu \rangle = f_{\langle \ue P^{-1}|} (\uc) -f_{\langle \ue P^{-1}|} (\uu) \\
- \phi'_{\langle \ue|, \langle \uk|}( \uc|\uw, \OO) &\longrightarrow 2 \pi \ii f_{\langle \ue P^{-1}| }(\uu) \phi_{\langle \uk P^{-1}|} ( \uc |\uv+\uu, \tO) - \phi'_{\langle \ue P^{-1}| , \langle \uk P^{-1}|} ( \uc |\uv+\uu, \tO). \label{eq:phi'Transf}
\end{align}
\begin{proof}
From the previous two transformation rules it follows that:
\begin{align}
\phi_{\langle \ue|, \langle \uk|}^{(\beta, p_i)} ( \uc|\uw, \OO) \longrightarrow & N_{\uk}^{(\beta,p_i)} C^{(\beta,p_i)} \left [ \left ( f_{\langle \ue P^{-1}|}(\uc) - \cancel{f_{\langle \ue P^{-1}|}(\uu)} \right ) \phi_{\langle \uk P^{-1}|}( \uc|\uv + \uu , \tO)\right . \nonumber \\
& \left . + \cancel{f_{\langle \ue P^{-1}|}(\uu) \phi_{\langle \uk P^{-1} |}( \uc|\uv + \uu , \tO) } 
- \frac{1}{2 \pi \ii}  \phi'_{\langle \ue P^{-1}| , \langle \uk P^{-1}|}( \uc |\uv + \uu , \tO)
\right ] \\
& = N_{\uk}^{(\beta,p_i)} C^{(\beta,p_i)} \left [ f_{\langle \ue P^{-1}|}(\uc) \phi_{\langle \uk P^{-1}|}(\uc |\uv + \uu , \tO) - \frac{1}{2 \pi \ii} \phi'_{\langle \ue P^{-1}|, \langle \uk P^{-1}|} (\uc| \uv + \uu , \tO) \right] \\
& =  \phi_{\langle \ue P^{-1}|, \langle \uk P^{-1}|}^{(\beta, p_i)} ( \uc|\uv + \uu , \tO).
\end{align}
\end{proof}
\end{prop}

\begin{prop}
As an aside we report the derivation of Eq.~\ref{eq:phi'Transf}:
\begin{align}
F \left [ - \phi'_{\langle \ue|, \langle \uk|}(\uc|\uw, \OO) \right ] = F \left [- e^{\ii \pi \uc \cdot \OO \uc} \, e^{-2 \pi \ii \uw \cdot \uk} \, D_{\langle \ue|}(\uk - \OO \uc) \ratthe{\uw}{\uzz}(\uk - \OO \uc| \OO) \right ]. \label{eq:phi'Transf2}
\end{align}
\begin{proof}
\begin{align}
\textrm{Eq. (\ref{eq:phi'Transf2})} = & - \left [ e^{\ii \pi \uc \cdot \tO \uc} \, e^{\ii \pi   \uu \cdot \tO \uu} \, e^{- 2 \pi \ii \uu \cdot \tO \uc}\right ]
\times \left [ e^{- 2 \pi \ii \uv \cdot {^tP^{-1}}\uk} e^{- \ii \pi  \uu \cdot \tO \uu} \, D_{\langle \ue P^{-1}|}(\utz) \left [ e^{-2 \pi \ii \uu \cdot \utz} \ratthe{\uv + \uu}{\uzz}(\utz|\tO) \right] \right ] \nonumber \\
= & - e^{\ii \pi \uc \cdot \tO \uc} \, \cancel{e^{-2 \pi \ii \uu \cdot \tO \uc}}\, e^{- 2 \pi \ii \uv \cdot {^tP^{-1}} \uk} \left [ - 2 \pi \ii \langle \ue P^{-1}| \uu \rangle \, e^{-2 \pi \ii \uu \cdot ({^tP^{-1}}\uk \cancel{-\tO \uc})} \ratthe{\uv + \uu}{\uzz}(\utz |\tO) \right. \nonumber \\ 
& + \left . e^{-2 \pi \ii \uu \cdot ({^tP^{-1}}\uk \cancel{-\tO \uc})} \, D_{\langle \ue P^{-1}|}(\utz) \ratthe{\uv + \uu}{\uzz}(\utz |\tO) \right ], \label{eq:phi'Transf3}
\end{align}
where in going from the first to the second line of Eq.~\ref{eq:phi'Transf3} we have simply used the chain rule for the derivative and canceled opposite terms as appropriate.
The last line of Eq.~\ref{eq:phi'Transf3} can be immediately recast as in Eq.~\ref{eq:phi'Transf}, by applying the appropriate definitions.
\end{proof}
\end{prop}

In order to obtain the transformed $d$-orbital states we need to derive the relevant transformations for the second derivative $\phi''$:
\begin{prop}
\begin{align}
F \left [ \phi''_{\langle \ue_1| \langle \ue_2 | , \langle \uk|} ( \uc| \uw, \OO) \right ] &=  \phi''_{\langle \ue_1 P^{-1}|\langle \ue_2 P^{-1}|, \langle \uk P^{-1}|}(\uc |\uv + \uu , \tO) 
- (2 \pi \ii) f_{\langle \ue_1 P^{-1}|}(\uu) \phi'_{\langle \ue_2 P^{-1} |,\langle \uk P^{-1}|}( \uc |\uv + \uu , \tO)  \nonumber \\
& - 2 \pi \ii f_{\langle \ue_2 P^{-1}|}(\uu) \phi'_{\langle \ue_1 P^{-1}| , \langle \uk P^{-1}|}(\uc|\uv + \uu , \tO)
+ (2 \pi \ii)^2 f_{\langle \ue_1 P^{-1}|}(\uu) f_{\langle \ue_2 P^{-1}|}(\uu) \phi_{\langle \uk P^{-1}|}( \uc|\uv + \uu , \tO). \label{eq:Transf_phi2}
\end{align}
\begin{proof}
Let's start by recalling the expression for $\phi''$:
\begin{align}
\phi''_{\langle \ue_1 | \langle \ue_2|, \uk}(\uw, \uc| \OO) & = \DD (\uk - \OO \uc) \left[ e^{\ii \pi \uc \cdot \OO \uc} \, e^{- 2 \pi \ii   \uw \cdot \uk} \, \ratthe{\uw}{\uzz}(\uk - \OO \uc| \OO) \right] \nonumber \\
& = e^{\ii \pi \uc \cdot \OO \uc} \, e^{- 2 \pi \ii \uw \cdot \uk} \DD (\uk - \OO \uc) \left[ \ratthe{\uw}{\uzz}(\uk - \OO \uc| \OO) \right].
\end{align}
The transformation laws applied to the individual terms produce the following expression:
\begin{align}
e^{\ii \pi \uc \cdot \tO \uc} \cancel{e^{\ii \pi   \uu \cdot \tO \uu}} \, e^{-2 \pi \ii \uu \cdot \tO \uc} \, e^{- 2 \pi \ii \uv \cdot {^tP}^{-1}\uk} \, \cancel{e^{- \ii \pi \uu \cdot \tO \uu}} \DDt (\utz) \left[ e^{- 2 \pi \ii \uu \cdot \utz} \, \ratthe{\uv + \uu}{\uzz}(\utz| \tO) \right], \label{eq:phi2Transf}
\end{align}
where the second derivative unpacks by the chain rule into:
\begin{align}
\textrm{Eq.(\ref{eq:phi2Transf})} & = e^{\ii \pi \uc \cdot \tO \uc} \, e^{- 2 \pi \ii \uu \cdot \tO \uc} \, e^{- 2 \pi \ii \uv \cdot {^tP}^{-1} \uk} 
\left \{ (2 \pi \ii )^2 \, \langle \ue_1 P^{-1}| \uu \rangle \langle \ue_2 P^{-1} | \uu \rangle \, e^{- 2 \pi \ii \uu \cdot \utz } \, \ratthe{\uv + \uu}{\uzz}(\utz | \tO) \right. \nonumber \\
 &  - 2 \pi \ii \langle \ue_1 P^{-1} | \uu \rangle \, e^{- 2 \pi \ii \uu \cdot \utz}\, D_{\langle \ue_2 P^{-1}|} \left[ \ratthe{\uv + \uu}{\uzz}(\utz | \tO) \right]
 - 2 \pi \ii \langle \ue_2 P^{-1} | \uu \rangle \, e^{- 2 \pi \ii \uu \cdot \utz} \, D_{\langle \ue_1 P^{-1}|} \left[ \ratthe{\uv + \uu}{\uzz}(\utz | \tO) \right]  \nonumber \\ 
 & \left. + e^{- 2 \pi \ii \uu \cdot \utz}\, \DDt \left[ \ratthe{\uv + \uu}{\uzz}(\utz | \tO ) \right] \right \}; \label{eq:phi2Transf2}
\end{align}

once the exponential prefactors are multiplied in, it's easy to recognise the familiar expression for the shifted $\phi$ function and its derivatives:
\begin{align}
\textrm{Eq. (\ref{eq:phi2Transf2})} & =e^{\ii \pi \uc \cdot \tO \uc} \,  e^{- 2 \pi \ii (\uv + \uu) \cdot {^tP}^{-1} \uk} \left \{  
(2 \pi \ii )^2 \, \langle \ue_1 P^{-1}| \uu \rangle \langle \ue_2 P^{-1} | \uu \rangle  \, \ratthe{\uv + \uu}{\uzz}(\utz | \tO) \right. \nonumber \\
 &  - 2 \pi \ii \langle \ue_1 P^{-1} | \uu \rangle \, D_{\langle \ue_2 P^{-1}|} \left[ \ratthe{\uv + \uu}{\uzz}(\utz | \tO) \right] 
 - 2 \pi \ii \langle \ue_2 P^{-1} | \uu \rangle \,  D_{\langle \ue_1 P^{-1}|} \left[ \ratthe{\uv + \uu}{\uzz}(\utz | \tO) \right]  \nonumber \\ 
 & \left. +  \DDt \left[ \ratthe{\uv +\uu}{\uzz}(\utz | \tO ) \right] \right \}.\label{eq:phi2Transf3}
\end{align}
By recognising the definition of the $\phi$ function and its derivatives in the expression above the statement of Eq.~\ref{eq:Transf_phi2} follows.
\end{proof}
\end{prop}

Now we have all the required entries to evaluate the transformed $d$-orbital state:
\begin{prop}[Transformation properties for $d$-orbital Bloch states]
\begin{align}
F \left [ \phi_{\langle \ue_1| \langle  \ue_2 |, \langle \uk |}^{(\beta, d_{ij})} (\uc|\uw , \OO)\right] &= N_{\uk}^{(\beta, d_{ij})} C^{(\beta, d_{ij})} \, \left \{ \vphantom{\frac{1}{2}} f_{\langle \ue_1 P^{-1} |} (\uc) f_{\langle \ue_2 P^{-1} |} (\uc)    \phi_{\langle \uk P^{-1}|}( \uc |\uv + \uu, \tO) \right.  \nonumber \\
& - \frac{1}{2 \pi \ii} f_{\langle \ue_1 P^{-1} |} (\uc) 
\phi'_{\langle \ue_2 P^{-1}|, \langle \uk P^{-1} |}( \uc|\uv + \uu, \tO) 
 - \frac{1}{2 \pi \ii} f_{\langle \ue_2 P^{-1} |} (\uc) 
\phi'_{\langle \ue_1 P^{-1}|, \langle \uk P^{-1}|}( \uc|\uv + \uu, \tO) \nonumber \\
& \left . + \frac{1}{(2 \pi \ii)^2} \phi''_{\langle \ue_1 P^{-1}| \langle \ue_2 P^{-1}|, \langle \uk P^{-1}|} ( \uc|\uv + \uu, \tO) \right \} = 
\phi_{\langle \ue_1 P^{-1}| \langle  \ue_2 P^{-1} |, \langle \uk P^{-1}|}^{(\beta, d_{ij})} ( \uc|\uv + \uu, \tO). \nonumber
\end{align}

\begin{proof}
To derive Eq.~\ref{eq:Transf_orb} for the case $I = d_{ij}$ we apply the transformation rules to the individual terms in Eq.~\ref{eq:d_state}:
\begin{align}
F \left [ \phi_{\langle \ue_1| \langle  \ue_2 |, \langle \uk|}^{(\beta, d_{ij})} (\uw, \uc| \OO)\right] & = 
N_{\uk}^{(\beta, d_{ij})} C^{(\beta, d_{ij})}
\left \{ \vphantom{\frac{1}{2}}\left ( f_{\langle \ue_1 P^{-1}|} (\uc) - f_{\langle \ue_1 P^{-1}|} (\uu) \right ) \left ( f_{\langle \ue_2 P^{-1}|} (\uc) - f_{\langle \ue_2 P^{-1}|} (\uu) \right )\right . \nonumber \\
& \times \phi_{\langle \uk P^{-1}|}( \uc |\uv + \uu , \tO) \nonumber \\
& - \frac{1}{2 \pi \ii} \left ( f_{\langle \ue_1 P^{-1}|} (\uc) - f_{\langle \ue_1 P^{-1}|} (\uu) \right ) \left [ -2 \pi \ii f_{\langle \ue_2 P^{-1}|} (\uu) \phi_{\langle \uk P^{-1}|}( \uc |\uv + \uu , \tO) \right . \nonumber \\
&\left . + \phi'_{\langle \ue_2 P^{-1}|, \langle \uk P^{-1}|}(\uc|\uv + \uu , \tO)\right ] \nonumber \\
& - \frac{1}{2 \pi \ii} \left ( f_{\langle \ue_2 P^{-1}|} (\uc) - f_{\langle \ue_2 P^{-1}|} (\uu) \right ) \left [ -2 \pi \ii f_{\langle \ue_1 P^{-1}|} (\uu) \phi_{\langle \uk P^{-1}|}(\uc |\uv + \uu , \tO) \right . \nonumber \\
&\left . + \phi'_{\langle \ue_1 P^{-1}|, \langle \uk P^{-1}|}( \uc|\uv + \uu , \tO)\right ] \nonumber \\
& + \frac{1}{(2 \pi \ii)^2} 
\left [ \phi''_{\langle \ue_1P^{-1}| \langle \ue_2 P^{-1}|, \langle \uk P^{-1}|}( \uc|\uv + \uu , \tO) 
 \right . \nonumber \\
& - 2 \pi \ii f_{\langle \ue_2 P^{-1}|}(\uu) \phi'_{\langle \ue_1P^{-1}|, \langle \uk P^{-1}|}(\uc |\uv + \uu , \tO) \nonumber \\
& - 2 \pi \ii f_{\langle \ue_1 P^{-1}|}(\uu)  \phi'_{\langle \ue_2P^{-1}|, \langle \uk P^{-1}|}( \uc |\uv + \uu , \tO) \nonumber \\
& \left . \left . + (2 \pi \ii)^2 f_{\langle \ue_1 P^{-1}|}(\uu) f_{\langle \ue_2 P^{-1}|}(\uu)  \phi_{\langle \uk P^{-1}|}( \uc |\uv + \uu , \tO) \right ] \vphantom{\frac{1}{2}}\right \}.
\end{align}
By cancellation of the terms with opposite signs the expression above evaluates immediately as:
\begin{align*}
F \left [ \phi_{\langle \ue_1| \langle  \ue_2 |, \langle \uk |}^{(\beta, d_{ij})} (\uc|\uw, \OO)\right] & =
\phi_{\langle \ue_1 P^{-1}| \langle  \ue_2 P^{-1} |, \langle  \uk P^{-1} |}^{(\beta, d_{ij})} ( \uc|\uv + \uu, \tO). 
\end{align*}
\end{proof}
\end{prop}


\section{Proof of the inversion symmetry protection theorem }
\label{sec:AppxThm}
In this appendix the proof of the inversion symmetry protection theorem as stated in Ref. \onlinecite{Mumford1983} is reported.
Let us point out that the period matrix $\OO$ defines a lattice in $\C^g$ as $L_\OO = \Z^g + \OO \Z^g$, with $g = 3$ in our case being the genus of the Riemann surface, then with the symbol $\Xi_0 \subset \C^g/L_\OO$ we indicate the set  $\{ \uz \in \C^g/L_\OO \, : \, \vartheta(\uz|\OO) = 0 \}$.
With this notation the statement of part b of theorem 3.10 in Ref. \onlinecite{Mumford1983} is:
\begin{prop}
The map $s = \begin{pmatrix}
q \\ p
\end{pmatrix} \mapsto \{ \textrm{zeroes of }\ratthe{q}{p}(\uz|\OO)\} \subset \C^g /L_\OO$ is an isomorphism between $\frac{1}{2} \Z^{2g}/\Z^{2g}$ and the set of translates $\ue + \Xi_0$ which are symmetric under inversion $\uz \rightarrow -\uz$.
\begin{proof}
We begin by noting that $\vartheta(\uz|\OO) = \vartheta(-\uz | \OO)$ (as it is immediately evident from the definition \ref{eq:theta}), so $\Xi_0$ is itself symmetric under inversion. 
Now, the set defined by $\ratthe{q}{p}(\uz | \OO)=0$ is the translate of $\Xi_0$ by $\ue = p + \OO q $ and if both $p,q \in \frac{1}{2}\Z^g/\Z^g$, then $p + \OO q \in \C^g/L_\OO$ is of order two and the translate of a symmetric element of order two is symmetric.
In fact, a translate of a symmetric subset $\Xi_0$ by an element $\uv \in \C^g /L_\OO$ is also symmetric under inversion if and only if it is invariant by a translation by $2 \uv$.
Therefore every element $\ue \in \frac{1}{2} \Z^g/\Z^g$ symmetric under inversion can be written in terms of the $p,q $ introduced previously and the map is onto.

Next, to prove that the map is also injective, we need to check that $\Xi_0 \neq \Xi_0 + \ue $ for all $\ue \neq \uzz$.
Let us assume that $\ue \in \C^g$ satisfies $\vartheta(\uz | \OO) =0 \Leftrightarrow \vartheta(\uz + \ue | \OO) = 0$, then we must show that $\ue \in L_\OO$.
Consider the ratio
\begin{align*}
\frac{\vartheta(\uz + \ue | \OO)}{\vartheta(\uz | \OO)}
\end{align*}
where the zeroes of numerator and denominator cancel each other out over the whole domain, the ratio is thus a non-zero holomorphic function on $\C^g$ and thus we can take the function:
\begin{align*}
f(\uz) = \ln \left( \frac{\vartheta(\uz + \ue | \OO)}{\vartheta(\uz | \OO)}\right) \Rightarrow \vartheta(\uz + \ue | \OO ) = e^{f(\uz)} \, \vartheta(\uz | \OO)
\end{align*}
With the replacement $\uz \mapsto \uz + \OO \un + \um$ and the use of the functional equation in Proposition \ref{prop:quasiperiod}-2. for the $\vartheta -$function on both sides, it follows that 
\begin{align*}
e^{-\ii \pi \un \cdot \OO \un} e^{-2\pi \ii \un \cdot (\uz + \ue)} e^{f(\uz)} \vartheta(\uz | \OO) = e^{f(\uz + \OO \un + \um)} e^{-\ii \pi \un \cdot \OO \un} e^{-2 \pi \ii \un \cdot \uz} \vartheta(\uz | \OO)
\end{align*}
whence
\begin{align}
f(\uz + \OO \un + \um) - f(\uz) \equiv -2 \pi \ii \, \un \cdot \ue 
\mod(2 \pi \ii ) \; \forall \uz \in \C^g
\label{eq:fz}
\end{align}
and therefore $\frac{\partial f}{\partial z_i}$ is invariant under $\uz \mapsto \uz + \OO \un + \um$, \textit{i.e.} it is an holomorphic function on $\C^g /L_\OO$.
But then, since $f$ has no poles, $\frac{\partial f}{\partial z_i}$ must be a constant and therefore $f$ is a linear function. 
By replacing $f(\uz) = c_0 + 2 \pi \ii \underline{c} \cdot \uz$ in Eq.~\ref{eq:fz} we get:
$c_0 + 2 \pi \ii \underline{c} \cdot (\uz + \OO \un + \um) - c_0 - 2 \pi \ii \underline{c}\cdot \uz \equiv -2 \pi \ii \un \cdot \ue \mod (2 \pi \ii)$ from which follows that $\underline{c} \cdot \um$ is an integer (and hence $\underline{c} \in \Z^g$).
This conclusion, together with the expression $\underline{c} \cdot \OO \un \equiv - \un \cdot \ue  \mod (1)$, allows us to write $\ue$ as $\ue = - \OO \underline{c} + \uv$, where $\uv \in \Z^g$ and therefore $\ue \in L_\OO$.
Hence there is a one-to-one correspondence between the zeroes of the $\vartheta -$functions with rational characteristics protected by the inversion symmetry and the Wyckoff positions and high symmetry wavevectors with half-integer coordinates.
	In the main text we have denoted the translated sets $\Xi_0 + \ue$ with the symbol $\Xi_s$, exploiting this one-to-one correspondence.
\end{proof}
\end{prop}


\section{Character tables for the Irreducible representations}
\label{sec:AppxCC}
In this Appendix we report for completeness the character tables in order to completely specify the Irreps of the two space groups considered. 
	The Irreps label correspond to those employed in the manuscript and the conjugacy classes are identified by a representative element and by the class exponent (Exp).
\subsection{Space group no. 207}

\begin{table}[h]
\renewcommand{\arraystretch}{1.2}
\centering
\begin{tabular}{c|c c c c c|c|c}
\hline
 $\mathsf{Stab}(\uw_a)$ &  &  & $\chi$ &  & & Exp & $\mathfrak{Cl}$ \\
\hline
$a_1$ & 1 &  1 &  1 &  1 &  1 & 1 & $(xyz|\uzz)$ \\
$a_2$ & 1 &  1 & -1 &  1 & -1 & 2 & $(x \oy \oz |\uzz )$\\
$e_3$ & 2 &  2 &  0 & -1 &  0 & 2 & $(z \oy x |\uzz )$\\
$t_4$ & 3 & -1 & -1 &  0 &  1 & 3 & $(y z x| \uzz)$\\
$t_5$ & 3 & -1 &  1 &  0 & -1 & 4 & $(x z \oy | \uzz)$\\
\hline
\end{tabular}
\caption{character table for the stabiliser of Wyckoff position $a$}
\label{tbl:StabWa_SG207}
\end{table}

\begin{table}[h]
\renewcommand{\arraystretch}{1.2}
\centering
\begin{tabular}{c|c c c c c|c|c}
\hline
$G_\Gamma$	&  &   & $\chi$  &  &  & Exp & $\mathfrak{Cl}$ \\
\hline
$a_1$& 1 &  1 &  1 &  1 &  1 & 1 & $(xyz|\uzz)$\\
$a_2$& 1 &  1 & -1 &  1 & -1 & 2 & $(x \oy \oz |\uzz )$\\
$e_3$& 2 &  2 &  0 & -1 &  0 & 2 & $(z \oy x |\uzz )$\\
$t_4$& 3 & -1 & -1 &  0 &  1 & 3 & $(y z x| \uzz)$\\
$t_5$& 3 & -1 &  1 &  0 & -1 & 4 & $(x z \oy | \uzz)$ \\
\hline
\end{tabular}
\caption{character table for $G_\Gamma \cong \mathsf{Sym}(4)$}
\label{tbl:G_Gamma207}
\end{table}

\begin{table}[h]
\renewcommand{\arraystretch}{1.2}
\centering
\begin{tabular}{c|c c c c c c c c c c|c|c}
\hline
$G_X$ & &  &  &  & $\chi$ &  &  &  &  &  &Exp & $\mathfrak{Cl}$\\
\hline
$a_1$ & 1  & 1  & 1  & 1  & 1  & 1  & 1  & 1  & 1  & 1  & 1 & $(x y z | \uzz)$\\

$a_2$ & 1  & -1  & 1  & -1  & 1  & -1  & -1  & 1  & 1  & -1 & 4 & $(z y \ox | \uzz)$\\

$a_3$ & 1  & -1  & 1  & -1  & 1  & 1  & 1  & -1  & -1  & 1  & 2 & $(\ox y \oz |\uzz)$\\

$a_4$ & 1  & -1  & 1  & 1  & -1  & -1  & -1  & 1  & -1  & 1 & 2 & $(x \oy \oz | \uzz)$\\

$a_5$ & 1  & -1  & 1  & 1  & -1  & 1  & 1  & -1  & 1  & -1  & 2 & $(z \oy x | \uzz)$\\

$a_6$ & 1  & 1  & 1  & -1  & -1  & -1  & -1  & -1  & 1  & 1  & 2 & $(x y z | 0, 1 , 0)$\\

$a_7$ & 1  & 1  & 1  & -1  & -1  & 1  & 1  & 1  & -1  & -1  & 2 & $(\ox y \oz |0, 1 , 0 )$\\

$a_8$ & 1  & 1  & 1  & 1  & 1  & -1  & -1  & -1  & -1  & -1  & 4 & $(z y \ox |0, 1 , 0)$\\

$e_9$ & 2  & 0  & -2  & 0  & 0  & -2  & 2  & 0  & 0  & 0  & 2 & $(x \oy \oz |0, 1 , 0)$\\

$e_{10}$ & 2  & 0  & -2  & 0  & 0  & 2  & -2  & 0  & 0  & 0  & 2 & $(z \oy x |0, 1 , 0 )$\\
\hline
\end{tabular}
\caption{character table for $G_X$}
\label{tbl:G_X207}
\end{table}

\begin{table}[h]
\renewcommand{\arraystretch}{1.2}
\centering
\begin{tabular}{c|c c c c c c c c c c|c|c}
\hline
$G_M$ &  &  &  &  & $\chi$ &  &  &  &  &  & Exp & $\mathfrak{Cl}$\\
\hline
$a_1$ & 1  & 1  & 1  & 1  & 1  & 1  & 1  & 1  & 1  & 1  & 1 & $(x y z | \uzz)$\\

$a_2$ & 1  & -1  & 1  & -1  & 1  & -1  & -1  & 1  & 1  & -1 & 4 & $(y \ox z |\uzz)$\\

$a_3$ & 1  & -1  & 1  & -1  & 1  & 1  & 1  & -1  & -1  & 1  & 2 & $(\ox \oy z|\uzz)$\\

$a_4$ & 1  & -1  & 1  & 1  & -1  & -1  & -1  & 1  & -1  & 1 & 2 & $(y x \oz | \uzz)$\\

$a_5$ & 1  & -1  & 1  & 1  & -1  & 1  & 1  & -1  & 1  & -1  & 2 & $(x \oy \oz | \uzz)$\\

$a_6$ & 1  & 1  & 1  & -1  & -1  & -1  & -1  & -1  & 1  & 1  & 2 & $(x y z | 0 , 1, 0)$\\

$a_7$ & 1  & 1  & 1  & -1  & -1  & 1  & 1  & 1  & -1  & -1  & 2 & $(\ox \oy z | 0, 1 , 0)$\\

$a_8$ & 1  & 1  & 1  & 1  & 1  & -1  & -1  & -1  & -1  & -1  & 4 & $(y \ox z|0, 1, 0)$\\

$e_9$ & 2  & 0  & -2  & 0  & 0  & -2  & 2  & 0  & 0  & 0  & 2 & $(y x \oz| 0, 1, 0)$\\

$e_{10}$ & 2  & 0  & -2  & 0  & 0  & 2  & -2  & 0  & 0  & 0  & 2 & $(x \oy \oz| 0, 1, 0)$\\
\hline
\end{tabular}
\caption{character table for $G_M$}
\label{tbl:G_M207}
\end{table}

\begin{table}[h]
\centering
\begin{tabular}{c|c c c c c c c c c c|c|c}
\hline
$G_R$ &  &  &  &  & $\chi$ &  &  &  &  & & Exp & $\mathfrak{Cl}$ \\
\hline
$a_1$ & 1  & 1  & 1  & 1  & 1  & 1  & 1  & 1  & 1  & 1 & 1 & $(x y z | \uzz)$ \\

$a_2$ & 1  & -1  & 1  & -1  & 1  & -1  & -1  & 1  & -1  & 1 & 2 & $(x y z | 0, 0, 1)$\\

$a_3$ & 1  & -1  & 1  & -1  & 1  & -1  & 1  & -1  & 1  & -1  & 2 & $(\ox \oy z|\uzz)$\\

$a_4$ & 1  & 1  & 1  & 1  & 1  & 1  & -1  & -1  & -1  & -1  & 2 & $(\ox \oy z|0, 0,1)$\\

$e_5$ & 2  & -2  & 2  & -2  & -1  & 1  & 0  & 0  & 0  & 0  & 3 & $(\oy z \ox|\uzz)$\\

$e_6$ & 2  & 2  & 2  & 2  & -1  & -1  & 0  & 0  & 0  & 0  & 6 &$(\oz \ox y| 0, 0, 1)$ \\

$t_7$ & 3  & -3  & -1  & 1  & 0  & 0  & -1  & 1  & 1  & -1 & 4 & $(\oy x z|\uzz)$ \\

$t_8$ & 3  & -3  & -1  & 1  & 0  & 0  & 1  & -1  & -1  & 1  & 4 & $(\oz y x | 0, 0, 1)$\\

$t_9$ & 3  & 3  & -1  & -1  & 0  & 0  & -1  & -1  & 1  & 1  & 2 & $(\ox z y| \uzz)$\\

$t_{10}$ & 3  & 3  & -1  & -1  & 0  & 0  & 1  & 1  & -1  & -1  & 2 & $(\ox z y | 0, 0, 1)$\\
\hline
\end{tabular}
\caption{character table for $G_R$}
\label{tbl:G_R207}
\end{table}

\clearpage

\pagebreak

\subsection{Space group no. 212}

\begin{table}[h!]
\centering
\begin{tabular}{c|c c c|c|c}
\hline
$\mathsf{Stab}(\uw_a)$ &  & $\chi$ &  & Exp & $\mathfrak{Cl}$\\
\hline
$a_1$ & 1 &  1 &  1 & 1 & $(xyz|\uzz)$\\
$a_2$ & 1 &  1 & -1 & 3 & $(zxy|\uzz)$\\
$e_3$ & 2 & -1 &  0 & 2 & $(\oz \oy \ox | \frac{1}{4}, \frac{1}{4}, \frac{1}{4})$\\
\hline
\end{tabular}
\caption{character table for the stabiliser of Wyckoff position $a$}
\label{tbl:StabWa_SG212}
\end{table}

\begin{table}[h]
\centering
\begin{tabular}{c|c c c|c|c}
\hline
$\mathsf{Stab}(\uw_c)$ &  & $\chi$ &  & Exp & $\mathfrak{Cl}$\\
\hline
$a_1$ & 1 & 1 & 1 & 1 & $(xyz | \uzz)$\\
$a_2$ & 1 & $-\frac{1}{2}+\ii \frac{\sqrt{3}}{2}$ & $-\frac{1}{2}-\ii \frac{\sqrt{3}}{2}$ & 3 & $(yzx | \uzz)$\\
$a_3$ & 1 & $-\frac{1}{2}-\ii \frac{\sqrt{3}}{2} $ & $-\frac{1}{2}+\ii \frac{\sqrt{3}}{2}$ & 3 & $(z x y | \uzz)$\\
\hline
\end{tabular}
\caption{character table for the stabiliser of Wyckoff position $c$}
\label{tbl:StabWc_SG212}
\end{table}

\begin{table}[h]
\renewcommand{\arraystretch}{1.2}
\centering
\begin{tabular}{c|c c c c c|c|c}
\hline
$G_\Gamma$	&  &   & $\chi$  &  &  & Exp & $\mathfrak{Cl}$ \\
\hline
$a_1$& 1 &  1 &  1 &  1 &  1 & 1 & $(xyz|\uzz)$\\
$a_2$& 1 &  1 & -1 & -1 &  1 & 2 & $(\ox \oy z | \frac{1}{2}, 0, \frac{1}{2})$\\
$e_3$& 2 &  2 &  0 &  0 & -1 & 4 & $(\oy x z | \frac{3}{4}, \frac{1}{4}, \frac{3}{4})$\\
$t_4$& 3 & -1 & -1 &  1 &  0 & 2 & $(y x \oz | \frac{1}{4}, \frac{3}{4}, \frac{3}{4})$\\
$t_5$& 3 & -1 &  1 & -1 &  0 & 3 & $(y z x | \uzz)$ \\
\hline
\end{tabular}
\caption{character table for $G_\Gamma \cong \mathsf{Sym}(4)$}
\label{tbl:G_Gamma212}
\end{table}

\begin{table}[h]
\renewcommand{\arraystretch}{1.2}
\centering
\begin{tabular}{c|c c c c c c c|c|c}
\hline
$G_X$	&  &  &  & $\chi$ &  &  &  & Exp & $\mathfrak{Cl}$ \\
\hline
$a_1$& 1 &  1 &  1 &  1 &  1 &  1 &  1  & 1 & $(xyz|\uzz)$\\
$a_2$& 1 & -1 & -1 &  1 &  1 & -1 &  1  & 8 & $(z y \ox| \frac{1}{4}, \frac{3}{4}, \frac{3}{4})$\\
$a_3$& 1 & -1 & -1 &  1 &  1 &  1 & -1  & 8 & $(\oz y x | \frac{3}{4}, \frac{1}{4}, \frac{3}{4})$\\
$a_4$& 1 &  1 &  1 &  1 &  1 & -1 & -1  & 4 & $(\ox y \oz | 0, \frac{1}{2}, \frac{1}{2})$\\
$e_5$& 2 & -$\sqrt{2}$   &  $\sqrt{2}$  &  0 & -2 &  0 &  0  & 2 & $(x y z| 0, \overline{1} , 0)$ \\
$e_6$& 2 &  0 &  0 & -2 &  2 &  0 &  0  & 2 & $(z \oy x| \frac{3}{4}, \frac{3}{4}, \frac{1}{4})$ \\
$e_7$& 2 & $\sqrt{2}$   &  -$\sqrt{2}$  &  0 & -2 &  0 &  0  & 2 & $(x \oy \oz| \frac{1}{2}, \frac{1}{2} , 0)$ \\
\hline
\end{tabular}
\caption{character table for $G_X \cong \mathsf{D}_{16}$}
\label{tbl:G_X212}
\end{table}

\begin{table}
\renewcommand{\arraystretch}{1.2}
\centering
\begin{tabular}{c|c c c c c c c c c c|c|c}
\hline
$G_M$&  &  &  &  & $\chi$ &  &  &  &  &  & Exp &$\mathfrak{Cl}$\\
\hline
$a_1$& 1 & 1 & 1 & 1 & 1 & 1 & 1 & 1 & 1 & 1 & 1 & $(xyz| \uzz)$\\

$a_2$& 1 & -$\ii$ & -1 & 1 &  $\ii$ & -1 & -$\ii$ & -1 & 1 &  $\ii$ & 4 & $(\oy x z | \frac{3}{4}, \frac{1}{4}, \frac{3}{4})$\\

$a_3$& 1 & -$\ii$ & -1 & 1 &  $\ii$ & -1 &  $\ii$ & 1 & -1 & -$\ii$ & 2 & $(x y z| 1, 0, \overline{1})$\\

$a_4$& 1 & 1 & 1 & 1 & 1 & 1 & -1 & -1 & -1 & -1 & 2 & $(\ox \oy z|\frac{1}{2}, 0, \frac{1}{2})$\\

$a_5$& 1 &  $\ii$ & -1 & 1 & -$\ii$ & -1 & -$\ii$ & 1 & -1 &  $\ii$ & 4 & $(y \ox z|\frac{3}{4}, \overline{\frac{1}{4}}, \overline{\frac{3}{4}})$\\

$a_6$& 1 &  $\ii$ & -1 & 1 & -$\ii$ & -1 &  $\ii$ & -1 & 1 & -$\ii$ & 2 &$(\ox \oy z| \overline{\frac{1}{2}}, 0 , \overline{\frac{1}{2}})$\\

$a_7$& 1 & -1 & 1 & 1 & -1 & 1 & 1 & -1 & -1 & 1 & 4 &$(\ox y \oz|0,\frac{1}{2}, \frac{1}{2})$\\

$a_8$& 1 & -1 & 1 & 1 & -1 & 1 & -1 & 1 & 1 & -1 & 2 &$(\oy \ox \oz| \frac{1}{4},\frac{1}{4},\frac{1}{4})$\\

$e_9$& 2 & 0 & 2 & -2 & 0 & -2 & 0 & 0 & 0 & 0 & 2 &$(y x \oz|\frac{1}{4},\frac{3}{4},\frac{3}{4})$\\

$e_{10}$& 2 & 0 & -2 & -2 & 0 & 2 & 0 & 0 & 0 & 0 & 4 & $(x \oy \oz| \frac{1}{2}, \frac{1}{2}, 0)$\\
\hline
\end{tabular}
\caption{character table for $G_M \cong (\Z_2 \times \Z_2) \rtimes \Z_4 $}
\label{tbl:G_M212}
\end{table}

\begin{table}
\renewcommand{\arraystretch}{1.2}
\centering
\begin{tabular}{c|c c c c c c c c|c|c}
\hline
$G_R$ & &  &  & $\chi$ &  &  &  &  & Exp & $\mathfrak{Cl}$\\
\hline
$a_1$ & 1  & 1  & 1  & 1  & 1  & 1  & 1  & 1  & 1 & $(x y z | \uzz)$\\

$a_2$ & 1  & 1  & -1 & 1  & -1 & -1 & 1  & 1  & 6 & $(\oy \oz x|\frac{1}{2}, 0 , \frac{1}{2})$ \\

$e_3$ & 2 & 1  &  -$\sqrt{2} \ii$ & 0  &  $\sqrt{2} \ii$ & 0  & -1 & -2 & 8 & $(y \ox z| \frac{3}{4}, \frac{3}{4}, \frac{1}{4})$ \\

$e_4$ & 2 & 1  &  $\sqrt{2} \ii$ & 0  & -$\sqrt{2} \ii$ & 0  & -1 & -2 & 4 & $(\ox \oy z| \frac{1}{2}, 0 , \frac{1}{2})$\\

$e_5$ & 2 & -1 & 0  & 2 & 0  & 0  & -1 & 2 & 8 & $(\oy x z| \frac{3}{4}, \frac{1}{4}, \frac{3}{4})$\\

$t_6$ & 3  & 0  & 1  & -1 & 1  & -1 & 0  & 3  & 2 & $(z \oy x|\frac{3}{4}, \frac{3}{4}, \frac{1}{4})$\\

$t_7$ & 3  & 0  & -1 & -1 & -1 & 1 & 0  & 3  & 3 &$(z \ox \oy | \frac{1}{2}, \frac{1}{2}, 0)$\\

$f_8$ & 4  & -1 & 0  & 0  & 0  & 0  & 1 & -4 & 2 & $(x y z| 0, 0 , 1)$\\
\hline
\end{tabular}
\caption{character table for $G_R \cong \mathsf{GL}_2(\mathbb{F}_3)$}
\label{tbl:G_R212}
\end{table}

\end{document}